\documentclass[aps,prl,preprint,groupedaddress]{revtex4-1}

\usepackage{graphicx}
\usepackage{color}

\begin{document}


\title{Influence of Ibuprofen on Phospholipid Membranes}


\author{Sebastian Jaksch}
\affiliation{Forschungszentrum J\"ulich GmbH, JCNS at Heinz Maier-Leibnitz Zentrum, Lichtenbergstra\ss e 1, D-85747 Garching}
\author{Frederik Lipfert}
\affiliation{Forschungszentrum J\"ulich GmbH, JCNS at Heinz Maier-Leibnitz Zentrum, Lichtenbergstra\ss e 1, D-85747 Garching}
\author{Alexandros Koutsioubas}
\affiliation{Forschungszentrum J\"ulich GmbH, JCNS at Heinz Maier-Leibnitz Zentrum, Lichtenbergstra\ss e 1, D-85747 Garching}
\author{Stefan Mattauch}
\affiliation{Forschungszentrum J\"ulich GmbH, JCNS at Heinz Maier-Leibnitz Zentrum, Lichtenbergstra\ss e 1, D-85747 Garching}
\author{Olaf Holderer}
\affiliation{Forschungszentrum J\"ulich GmbH, JCNS at Heinz Maier-Leibnitz Zentrum, Lichtenbergstra\ss e 1, D-85747 Garching}
\author{Oxana Ivanova}
\affiliation{Forschungszentrum J\"ulich GmbH, JCNS at Heinz Maier-Leibnitz Zentrum, Lichtenbergstra\ss e 1, D-85747 Garching}
\author{Samira Hertrich}
\affiliation{Ludwig-Maximilians-Universit\"at, Department f\"ur Physik und CeNS, Geschwister-Scholl-Platz 1, D-80539 M\"unchen}
\author{Stefan F. Fischer}
\affiliation{Ludwig-Maximilians-Universit\"at, Department f\"ur Physik und CeNS, Geschwister-Scholl-Platz 1, D-80539 M\"unchen}
\author{Bert Nickel}
\affiliation{Ludwig-Maximilians-Universit\"at, Department f\"ur Physik und CeNS, Geschwister-Scholl-Platz 1, D-80539 M\"unchen}
\author{Henrich Frielinghaus}
\affiliation{Forschungszentrum J\"ulich GmbH, JCNS at Heinz Maier-Leibnitz Zentrum, Lichtenbergstra\ss e 1, D-85747 Garching}

\date{\today}

\begin{abstract}
Basic understanding of biological membranes is of paramount importance as these membranes comprise the very building blocks of life itself. Cells depend in their function on a range of properties of the membrane, which are important for the stability and function of the cell, information and nutrient transport, waste disposal and finally the admission of drugs into the cell and also the deflection of bacteria and viruses.

We have investigated the influence of ibuprofen on the structure and dynamics of L-$\alpha$-phosphatidylcholine (SoyPC) membranes by means of grazing incidence small-angle neutron scattering (GISANS), neutron reflectometry and grazing incidence neutron spin echo spectroscopy (GINSES). From the results of these experiments we were able to determine that ibuprofen induces a two-step structuring behavior in the SoyPC films, where the structure evolves from the purely lamellar phase for pure SoyPC over a superposition of two hexagonal phases to a purely hexagonal phase at high concentrations. Additionally, introduction of ibuprofen stiffens the membranes. This behavior may be instrumental in explaining the toxic behavior of ibuprofen in long-term application.
\end{abstract}


\maketitle

\section{Introduction}
Phospholipid membranes are widely used as model systems for the study of the more complicated biological cell membranes. By these investigations information about the structure and behavior of these membranes are gained, which in turn are indispensable in today's medical and biological science. Aeffner et al. \cite{Aeffner2009, Aeffner2012} have reported stalk formation in lipid membranes for a variety of phospholipids. In that case the structural ordering in the membranes was induced by different relative humidities. Another possible way to induce ordering in phospholipid membranes is by electric fields as done by Tronin et al. \cite{Tronin2013}. Here however we want to maintain near-physiological conditions while inducing ordering into a lipid membrane. To do so, we use ibuprofen, which is known to decrease the elasticity of phospholipid membranes \cite{Boggara2010} and is moreover a common non-steroid anti-inflammatory (NSAID) drug with a wide range of possible applications, ranging from the treatment of cancer \cite{Thun2002}, Alzheimer's \cite{Weggen2001} and inflammations to the use as a painkiller. SoyPC is a phospholipid with two hydrocarbon tails that will facilitate the description of the membrane during the data evaluation process, if assumed as a pure hydrocarbon layer.

However, ibuprofen is also reported to be cytotoxic in oral long-term application \cite{Lichtenberger1995}, leading to sometimes-fatal ulcers and other gastrointestinal complications such as stomach bleeding. Investigating the influence of the ibuprofen concentration on structure formation within a phospholipid film may help elucidate the cause for this toxicity. Previous studies find the increased permeability of the cell membrane and thus the viability of the cell was linked to the NSAID content. \cite{Boggara2012, Tomisato2004}. Studies with chemically similar local anesthetics have been conducted by Malheiros et al. \cite{Malheiros}.

In the present system of ibuprofen/SoyPC we observe a structural evolution from lamellar over bi-hexagonal to single hexagonal lattices. A similar hexagonal near-surface structure in soft matter systems has also been reported for $C_m E_n$ surfactant/water systems \cite{Lang1998}. Additionally to these structural studies with grazing incidence neutron scattering (GISANS) and neutron reflectometry we also performed a kinetic study with grazing incidence neutron spin echo spectroscopy (GINSES). With this technique it is possible to detect kinetics of the film strictly perpendicular to the film surface on the nanometer scale.
\section{Experimental}
\subsection{Materials and Sample Preparation}
\begin{figure}
  \begin{minipage}[t]{0.66\textwidth}
    \raggedright{\large{(a)}}
  \includegraphics[width=\textwidth]{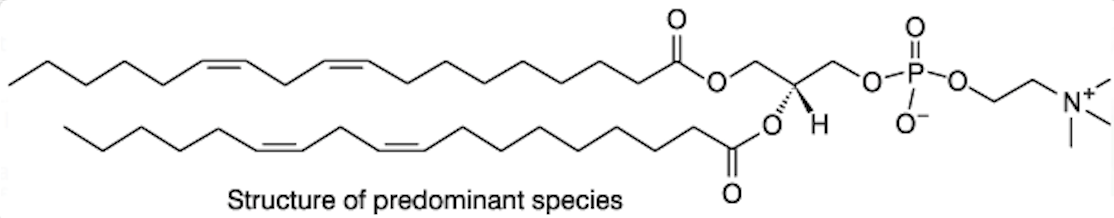} 
  \end{minipage}
  \begin{minipage}[t]{0.3\textwidth}
    \raggedright{\large{(b)}}
  \includegraphics[width=\textwidth]{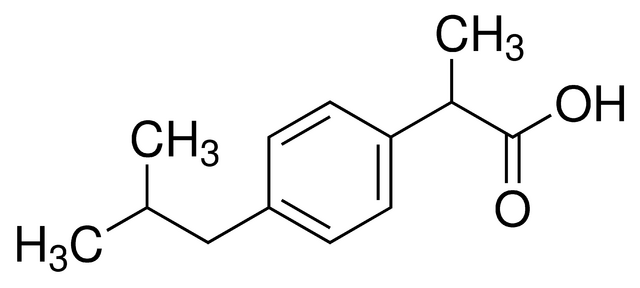} 
  \end{minipage}
   \caption{Structures of a) SoyPC \cite{SoyPC} and b) ibuprofen \cite{IbuSigma}. \label{fig:structures}}
 \end{figure}

The SoyPC was obtained in powder form from Avantilipids (Alabaster/AL, USA), the ibuprofen from Sigma Aldrich (M\"unchen, Germany), solvent was in all cases isopropanol pA (Roth, Karlsruhe, Germany), structures are given in fig.\,\ref{fig:structures}. Standard solutions of SoyPC in isopropanol were prepared with a molar ratio of 1.77\,mol\% between SoyPC and isopropanol. The mixing ratio was chosen in a way to ensure homogeneous mixing and easy handling during the preparation but has no discernible impact on the final sample as the sample is dried completely after the preparation. The resulting solutions were stirred for at least 20\,min. each. Subsequently the appropriate amounts of ibuprofen were added as given in table \ref{tab:mixing} and stirred again for at least 20\,min. which resulted in clear solutions.

 \begin{table}
 \caption{Mixing weights of SoyPC and ibuprofen\label{tab:mixing}}
 \begin{ruledtabular}
 \begin{tabular}{|c|c|c|c|c|c|c|c|}
				&	0 mol\%	&	13.6 mol\%	&	25.0 mol\%	&	34.5 mol\%	&	43.1 mol\%	&	50.2 mol\%	&	53.3 mol\%\\
ibuprofen / mg	&	0		&	138		&	290		&	458		&	658		&	679		&	3112\\
SoyPC / mg		&	2610		&	3229		&	3189		&	3189		&	3189		&	2479		&	10030\\
 \end{tabular}
 \end{ruledtabular}
 \end{table}

The silicon blocks (2x5x12\,$cm^3$) were prepared for deposition of these solutions by an RCA treatment \cite{Kern1970} after being cleaned in an ultrasonic bath until all optical impurities were removed. One side of the blocks was polished to a roughness of less than 5\,\AA. All solvents were obtained from Roth. The first cleaning bath  consisted of 280\,mL Millipore filtered and de-ionized water, 70\,mL H$_2$O$_2$ (30\%) and 70\,mL HCl (37\%). Treatment time was one hour at a temperature of 28$^\circ$C. The second bath was 280\,mL Millipore filtered and de-ionized water, 70\,mL H$_2$O$_2$ (30\%) and 70\,mL NH$_3$ (28\%). Treatment time was again one hour at 33$^\circ$C.

After cleaning of the silicon  12\,mL of the prepared solution were deposited on the blocks and dried at room temperature (22$^\circ$C) at a pressure of 250\,mbar. This pressure had to be maintained, otherwise superheating and bubbling of the film occurred. An o-ring in a custom made scaffold made sure the solution stayed on top of the silicon block and the silicon block was adjusted using a spirit level. After 24 hours of drying no remaining isopropanol could be detected either visually or by smell. This resulted in SoyPC layers of macroscopic dimensions (about 2\,mm in thickness). 

Immediately after preparation the coated silicon blocks were mounted into the sample cell. The sample cell was then filled with D$2$O (99.8\%) and mounted in the respective instrument. In the instruments, the sample cells were kept at 35$^\circ$C using a water thermostat. The sample cell allowed for visual inspection after filling and after performing the measurement. During this time no deterioration of the film coverage on the silicon block was detected.

\subsection{Sample Cell}
\begin{figure}
 \includegraphics[width=0.5\textwidth]{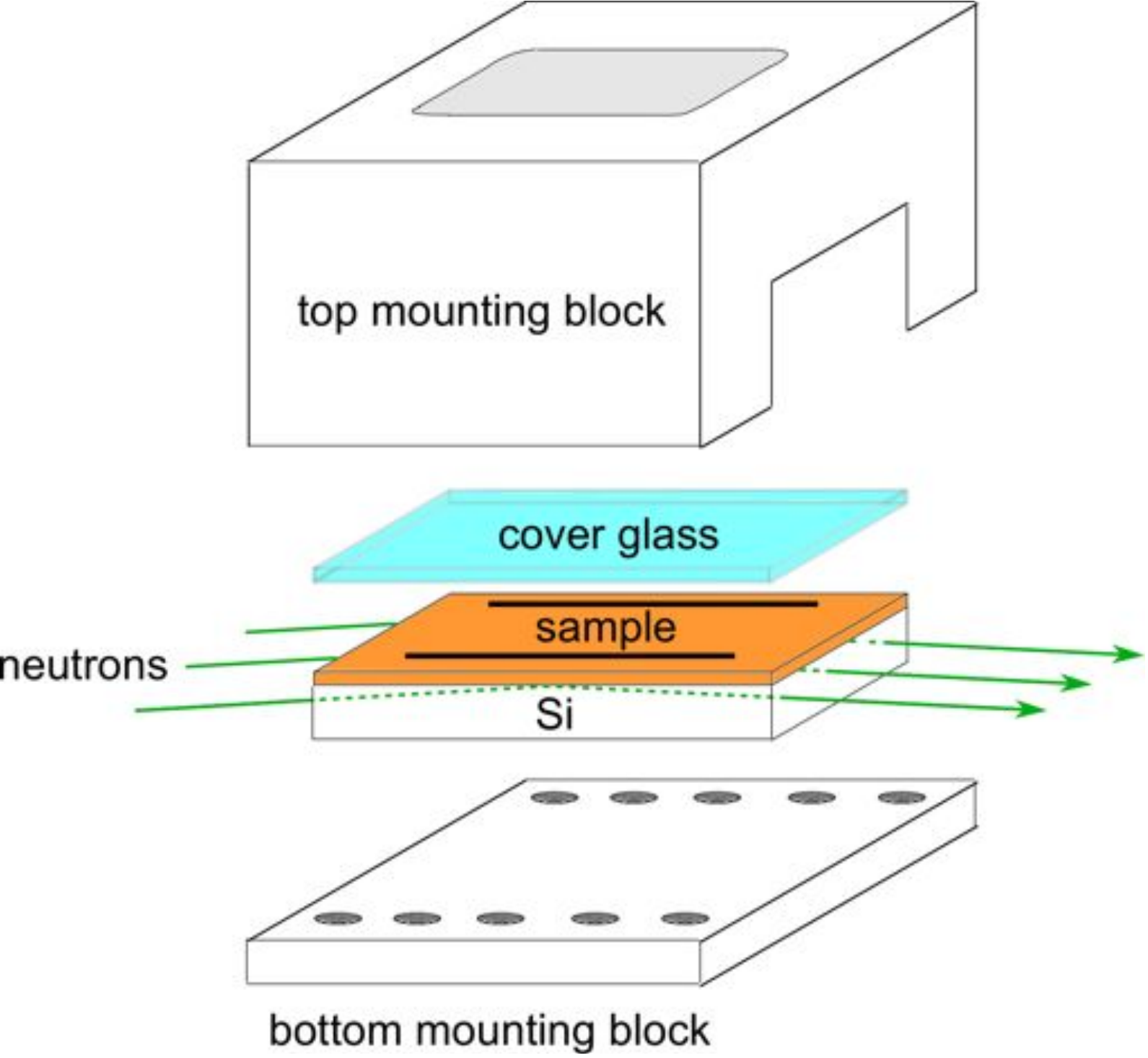}%
 \caption{(color online) Sketch of the sample cell used to mount the silicon block. The two black stripes on the sample are cadmium spacers to keep the cover glass from touching the sample. Additional o-rings (not shown) ensure the sample cell does not leak. Various drill holes (not shown) allow for attaching the sample cell to a water thermostat, as well as filling the space between sample and cover glass with D$_2$O.\label{fig:cell}}
 \end{figure}

The sample cell was designed to allow for GISANS, neutron reflectometry and GINSES consecutively, so the sample could be measured in all experiments in the same sample cell. A sketch of the sample cell is shown in fig. \ref{fig:cell}. Neutrons can enter the silicon block at the flat surface on the long side of the block (2x5\,$cm^2$). Due to the low scattering length density (SLD) of silicon it has a high absorption length, which is of the same order as the length of the block (52.7\% transmission at a wavelength of $\lambda=7$\,\AA\, along the long axis of the block). This geometry allows for a good control of the sample/silicon interface, where the scattering takes place, as opposed to the sample air interface where evaporation or scattering at the cover glass would take place. Also, this setup was designed to achieve a system that is oriented parallel to the surface of the silicon due to the hydrophobicity of the surface. At the same time neutrons are not unduly blocked by the silicon. Only in this setup it is possible to provide an initial orientation of the lipid layers along the block surface, so they can be investigated by reflectometry and GISANS. The hydrophilic surface of the silicon block keeps the lamellae aligned. Additionally to these reasons, at the solid/liquid interface there is no total external reflection as opposed to the air/liquid interface.

\subsection{Methods}
\subsubsection{GISANS}
Grazing Incidence Small-Angle Neutron Scattering (GISANS) was also performed at MARIA at MLZ, Garching, Germany. The wavelengths of the neutrons was set to $\lambda = 5$\,\AA\,at a wavelength spread of $\Delta\lambda/\lambda=0.1$.

\begin{figure}
\includegraphics[width=0.5\textwidth]{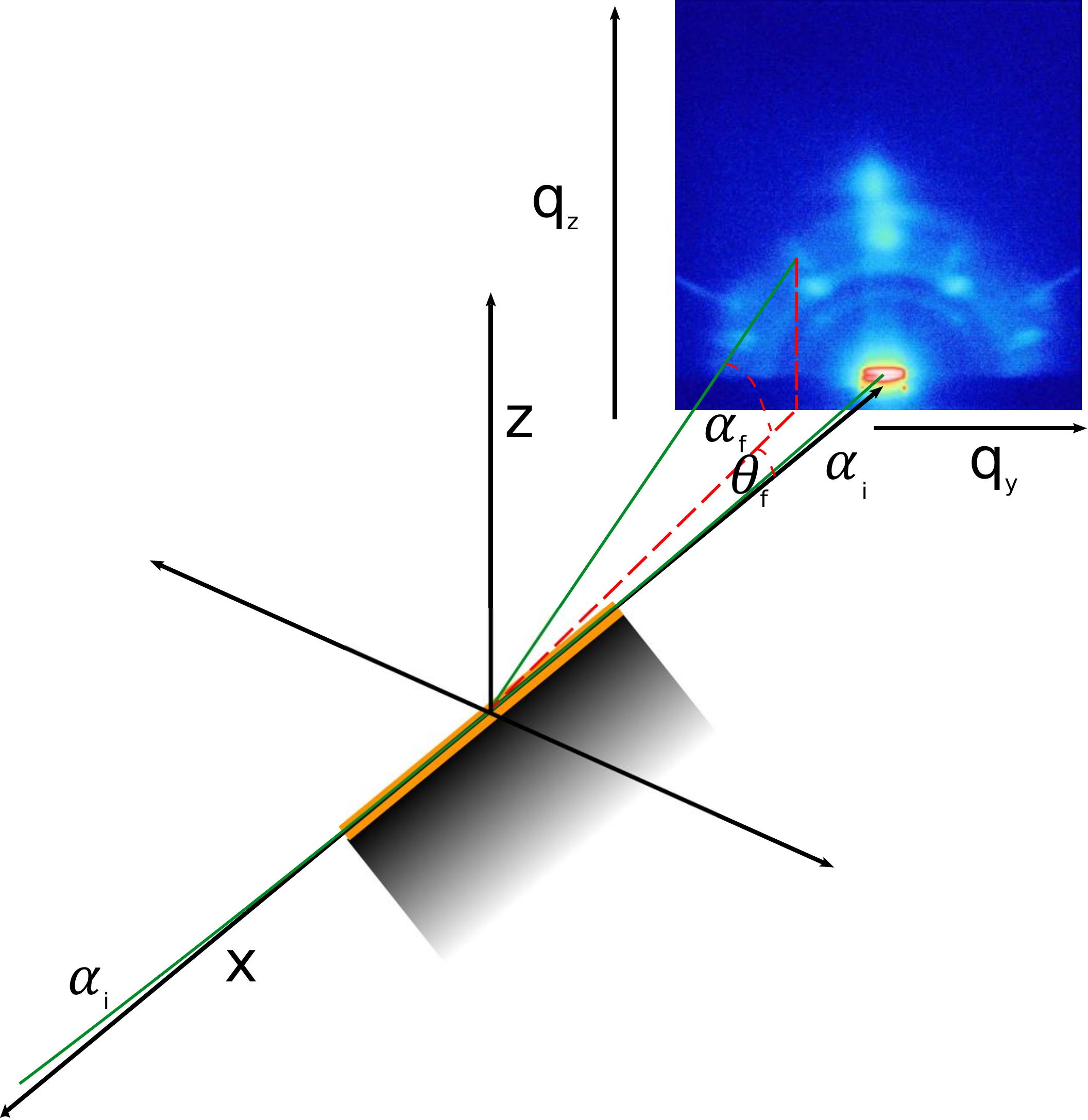}
 \caption{(color online) Geometry of a GISANS experiment. In case of reflectometry only intensity for $\alpha_i=\alpha_f$ is evaluated. The path of the netrons is in green. \label{fig:GISANS-geo}}
 \end{figure}

GISANS is a technique comparable to conventional Small-Angle Neutron Scattering (SANS) as the scattered intensity is the Fourier transform of the irradiated structure \cite{PMB}. However, instead of the beam impinging on the sample head-on (zero degree incident angle) the sample is irradiated under a shallow angle below the critical angle of total reflection. This way, instead of investigating the directly reflected beam under reflective conditions (incoming angle equal to outgoing angle) as reflectometry, scattered intensity over the complete detector is investigated (off-specular scattering). In GISANS measurements are mostly performed below the critical angle of total internal reflection $\alpha_c=\lambda\sqrt{\Delta\rho/\pi}$, where $\Delta\rho=\rho_{\mathrm{film}}-\rho_{\mathrm{substrate}}$ is the scattering length density contrast between the film and the substrate. At these conditions an evanescent wave with an exponentially decaying penetration depth of $\Lambda_{\mathrm{eva}}=\left[\mathrm{Re}\sqrt{4\pi\Delta\rho(1-\alpha_{in}^2/\alpha_c ^2)}\right]^{-1}$ is propagating into the sample, so information over the complete surface region down to the depth of the evanescent wave is averaged by the scattering \cite{Frielinghaus2012}. The geometry is shown in fig.\,\ref{fig:GISANS-geo}. In our case $\Lambda_{\mathrm{eva}}$ can be estimated to $\Lambda_{\mathrm{eva}}\approx350\mathrm{\AA}\gg D_{lam}\approx 50\mathrm{\AA}$ with $D_{lam}$ the thickness of the observed lamellae. The dominant layer signal is thus scattered from $\approx 7$ layers of the multilamellar system.

In the case investigated here we find two features we want to analyze: Distinct peaks and rings of near uniform intensity. The peaks can be described by classical crystallography and were indexed using TREOR90 \cite{Werner1985}. As the primitive tetragonal lattice that was found was not oriented parallel to the surface, the orientation was determined by comparison with simulated scattering images created by Xray-View 4.0 \cite{Phillips1995}. The peaks with a hexagonal symmetry can be explained by a hexagonal symmetry in the sample itself \cite{Lang1998}.

The rings can be described as diffuse Debye-Scherrer rings around the direct beam and the specular reflected beam \cite{Busch2007}. Since the thickness of these layers is already determined by reflectometry we will mainly concentrate on the relative intensity contribution to calculate the volume fractions of ordered and lamellar phases.

\subsubsection{Neutron Reflectometry}
Neutron reflectometry was performed at MARIA at MLZ, Garching, Germany. Reflectometry probes the sample composition on the nanometer scale perpendicular to the surface of the sample and is thus an ideal tool for the investigation of layered materials. A wavelength of $\lambda=10$\,\AA\,at a wavelength spread of $\Delta\lambda/\lambda = 0.1$ was used. Data acquisition time for each reflectometry point was 60\,s.

To evaluate the data the Parrat algorithm \cite{Parratt1954} was used. This algorithm describes the SLD distribution by describing it as a stack of discrete layers. The reflective properties of each layer $j$ in the multilayer stack can be described by the transition matrix $M_j$ \cite{Roe}.

\begin{equation}
M_j = \left(\begin{array}{cc}
\cos\phi_j & -(1/k_{zj})\sin\phi_j \\
k_{zj}\sin\phi_j & \cos\phi_j
\end{array}
\right).
\end{equation}

Here the phase difference $\phi_j$ is given by the incident angle $\theta$, the refractive index $n$, the wavelength of the neutrons $\lambda$ and the thickness of the $j$-th layer $t_j$ via

\begin{equation}
\phi_j=\frac{2\pi}{\lambda}n_j\sin\theta _j t_j=k_{zj}t_j.
\label{eq:phi}
\end{equation}

The matrix $M_j$ describes the amplitude of a wave propagating through layer $j$ to the layer boundary $(j,j+1)$ in relation to the behavior at the boundary $(j-1,j)$. As the amplitude and its derivation have to be continuous it is possible to construct a transition matrix $M$ for the whole stack of $N-1$ layers on the substrate, which is medium $N$ and infinitely thick (therefore giving boundary conditions of a zero amplitude):

\begin{equation}
M=\left(\begin{array}{cc}
m_{11} & m_{12} \\
m_{21} & m_{22}
\end{array}
\right) = M_{N-1}M_{N-2}M_{N-3}\cdots M_2 M_1.
\end{equation}

The reflective coefficient $R$ of  the $(0,1)$ interface is the given by

\begin{equation}
R=\frac{(k_{z0}k_{zN}m_{12}+m_{21})-i(k_{zN}m_{11}-k_{z0}m_{22})}{(k_{z0}k_{zN}m_{12}-m_{21})+i(k_{zN}m_{11}+k_{z0}m_{22})}.
\end{equation}

A set of matrices is found numerically, which minimizes the difference between the experimental data and the calculated reflected intensity. This allows to determine the scattering length density $\rho$ of each layer $j$ by

\begin{equation}
n=1-\frac{\lambda^2\rho}{2\pi}.
\label{eq:ref-index}
\end{equation}

These calculations have been implemented in the GenX Software \cite{Bjorck2007}.

\subsubsection{GINSES}
Grazing Incidence Neutron Spin-Echo Spectroscopy (GINSES) was developed at the \mbox{J-NSE} at the MLZ, Garching, Germany. The same sample cell and geometry as for the reflectometry and GISANS measurements was used. The wavelength was set to 8\,\AA, while the incoming angle was set to 0.21$^\circ$. The detector was placed at a $Q$-value of $Q=0.12$\AA$^{-1}$. This resulted in counting rates of $\approx 1$ cps. These low counting rates can be explained by the fact, that in contrast to conventional NSE experiments \cite{Monkenbusch1997, Holderer2008} the scattering volume only comprises the volume covered by the evanescent wave. However, this is only about 400\,\AA\, in thickness. An detailed description of the data analysis for the result can be found in Frielinghaus et al. \cite{Frielinghaus2012}. Due to the low countrates however here we limit the analysis to a qualitative interpretation.

\subsection{Results}
In this section we first present the results of the single GISANS, neutron reflectometry and GINSES separately. Afterwards the results will be compared and discussed in context among each other.

\subsubsection{GISANS}
\begin{figure}
  \begin{minipage}[b]{0.3\textwidth}
  \raggedright{\large{(a)}\hspace{1.5cm} 0\,mol\%}
 \includegraphics[width=\textwidth]{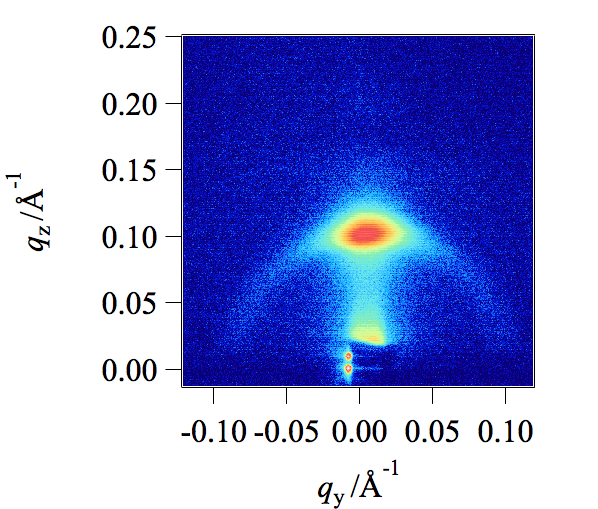} 
  \end{minipage}
  \begin{minipage}[b]{0.3\textwidth}
  \raggedright{\large{(b)}\hspace{1.225cm} 13.6\,mol\%}
  \includegraphics[width=\textwidth]{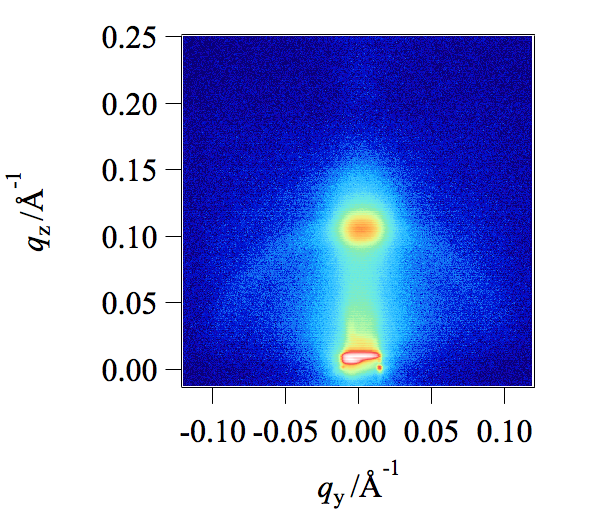} 
  \end{minipage}
  
  \begin{minipage}[b]{0.3\textwidth}
  \raggedright{\large{(c)}\hspace{1.225cm} 25.0\,mol\%}
  \includegraphics[width=\textwidth]{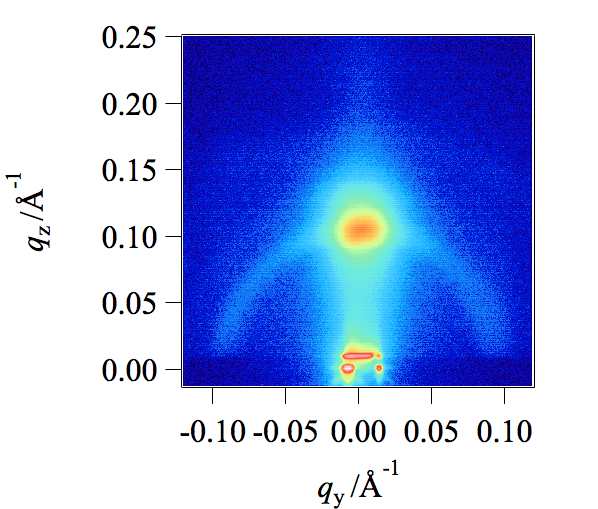} 
  \end{minipage}
  \begin{minipage}[b]{0.3\textwidth}
  \raggedright{\large{(d)}\hspace{1.225cm} 34.5\,mol\%}
  \includegraphics[width=\textwidth]{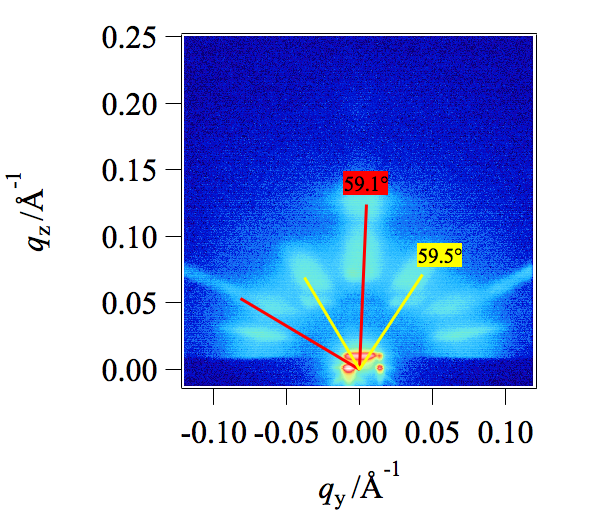} 
  \end{minipage}
  
    \begin{minipage}[b]{0.3\textwidth}
    \raggedright{\large{(e)}\hspace{1.225cm} 43.1\,mol\%}
  \includegraphics[width=\textwidth]{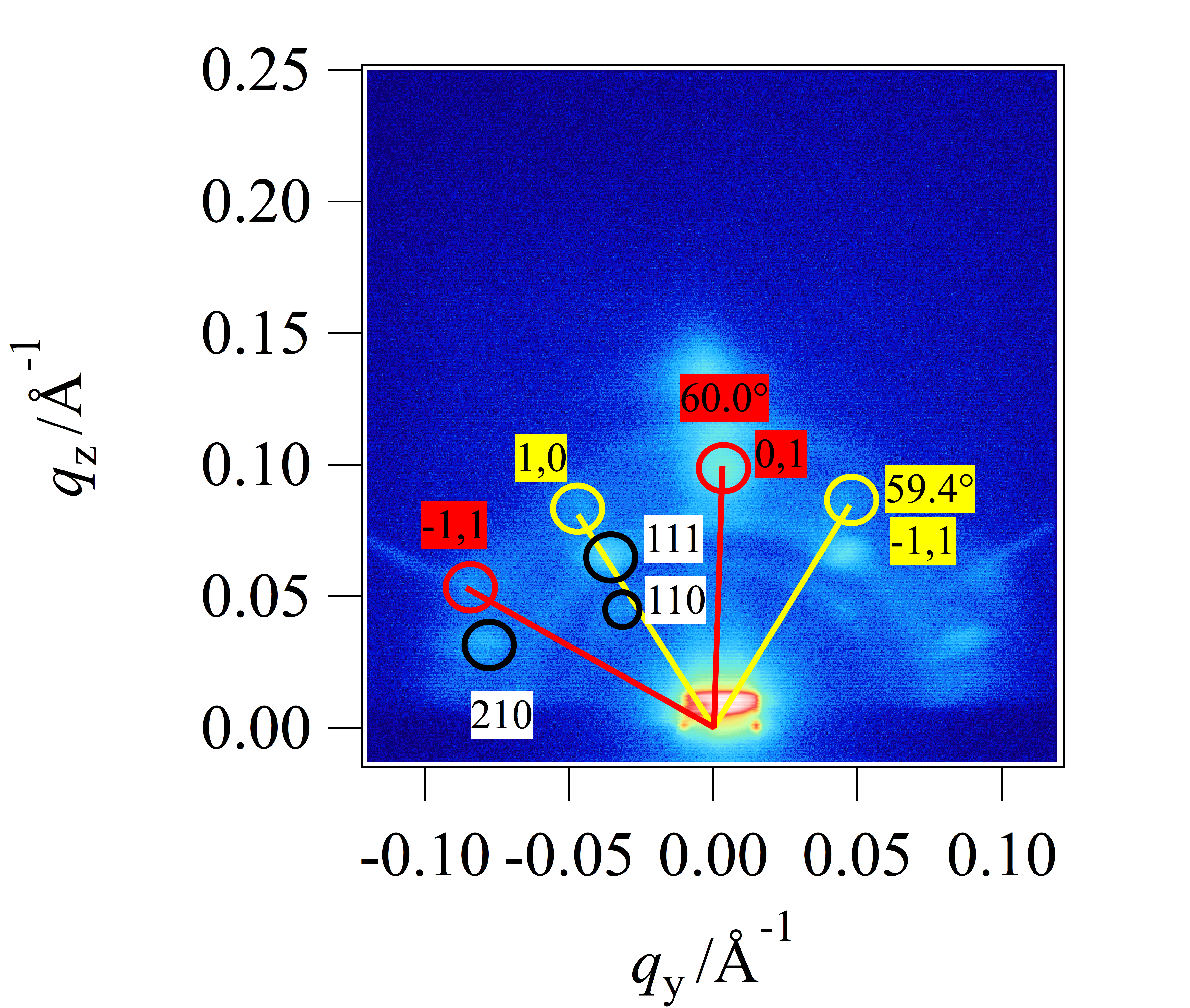} 
  \end{minipage}
  \begin{minipage}[b]{0.3\textwidth}
  \raggedright{\large{(f)}\hspace{1.225cm} 50.2\,mol\%}
  \includegraphics[width=\textwidth]{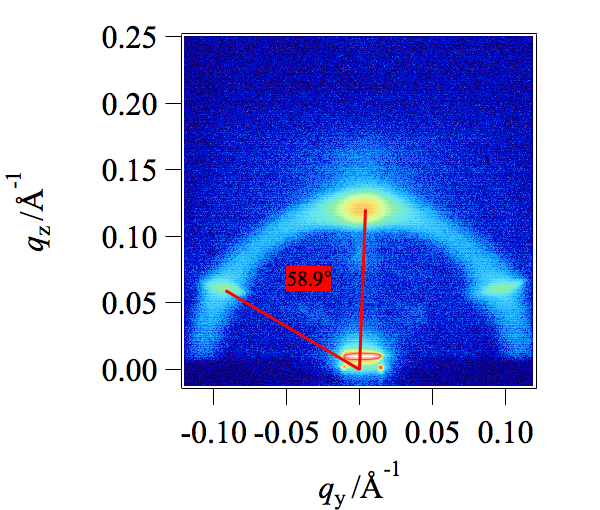} 
  \end{minipage}
  \begin{center}
    \begin{minipage}[b]{0.3\textwidth}
    \raggedright{\large{(g)}\hspace{1.225cm} 53.3\,mol\%}
  \includegraphics[width=\textwidth]{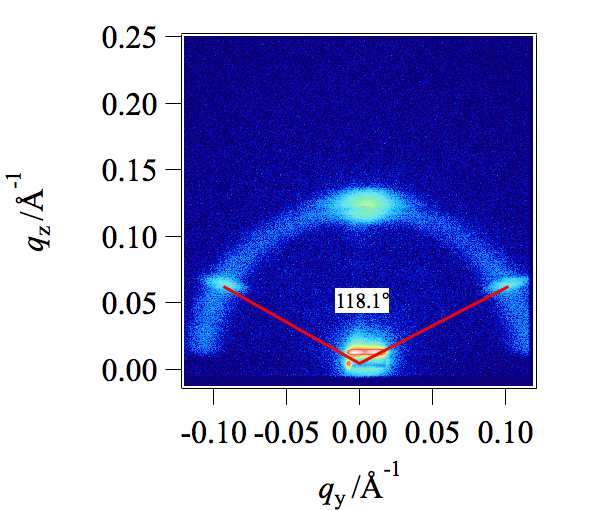} 
  \end{minipage}
  \end{center}
  \vspace{-0.8cm}
 \caption{(color) GISANS images at an incident angle of 0.2$^\circ$. Concentrations of ibuprofen are (a) 0\,mol\%, (b) 13.6\,mol\%, (c) 25.0\,mol\%, (d) 34.5\,mol\%, (e) 43.1\,mol\%, (f) 50.2\,mol\%, (g) 53.3\,mol\%. Scattering with a hexagonal symmetry is shown by yellow and red lines, colored labels show the angles between the respective lines. Indexed peaks are shown by red (hexagonal lattice with parallel axis to the surface), yellow (hexagonal lattice with perpendicular axis to the surface) and black (primitive tetragonal lattice) circles, where labels show the indexes. \label{fig:GISANS-images}}
\end{figure}

An overview over the scattering images obtained by the GISANS measurements at all investigated concentrations is shown in fig. \ref{fig:GISANS-images}. These images show a clear evolution from a lamellar based scattering, over a scattering where several different structures contribute to a hexagonal structure with additional disordered lamellae as can be seen by the persisting Debye-Scherrer Ring.

At low concentrations one single main maximum from lamellar scattering is visible. While hardly visible at 0\,mol\% a Debye-Scherrer ring starts to appear at 13.6 and 25.0\,mol\%, which is indicative of an increasing amount of still lamellar, yet disordered, scattering. This can be described as a powder of lamellar regions in the scattering volume. The second order peak can be found only in a linecut (see supporting information). It is three orders of magnitude smaller than the primary peak and thus not visible with the bare eye.

 \begin{figure}
 \includegraphics[width=0.8\textwidth]{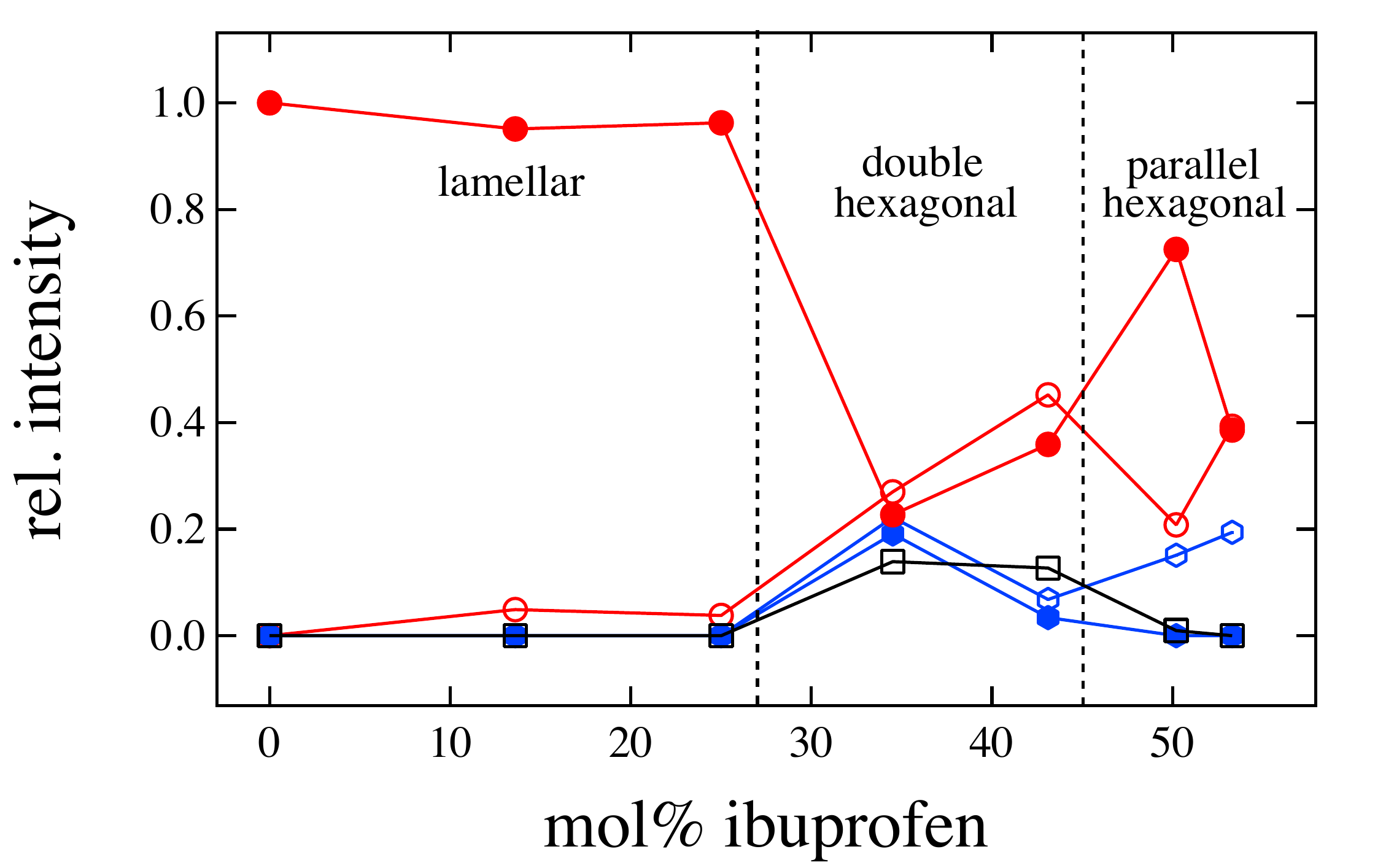}%
 \caption{(color online) Relative scattering contributions from lamellae parallel to the surface (red solid circles), disordered lamellae (open red circles), primitive tetragonal lattice (black open squares), hexagonal lattice with a parallel axis to the surface (open blue hexagons) and hexagonal lattice standing on an edge on the surface (blue solid hexagons). The vertical dashed lines mark the boundaries of the different regimes where lamellar scattering, scattering from two superimposed hexagonal lattices and scattering from a hexagonal lattice with an axis parallel to the substrate dominates.\label{fig:rel_intensity}}
 \end{figure}

Hexagonal structures start to emerge at 34.5\,mol\%. We find two simultaneously appearing hexagonal structures, where one exhibits a parallel axis to the substrate, while the other one is standing on an edge. There are also additional maxima, but they can only be indexed in the case of the 43.1\,mol\% GISANS image, because for all other concentrations the peaks from the hexagonal lattice are so strong they hide the exact location of these peaks. In the 43.1\,mol\% GISANS image these peaks were indexed as the (110), (111) and (210) peaks of a primitive tetragonal lattice. The lattice parameters are $a=74.6$\,\AA, $b=74.6$\,\AA\, and $c=64.9$\,\AA. All angles are $\alpha=\beta=\gamma=90^\circ$. The inclination of the unit cell was determined by comparison with simulated scattering images from XrayView 4.0 to be $77.3^\circ$. They are quite weak, which also means that the primitive tetragonal lattice only occupies a small volume fraction of the sample.

Only a single hexagonal lattice is retained at 50.2\,mol\% and above. In these instances the hexagonal lattice with an axis parallel to the substrate is still present.

The relative scattering contributions of the respective lattices, lamellae and Debye-Scherrer rings are shown in fig. \ref{fig:rel_intensity}. These relative intensities have been calculated using regions of interest (ROI) in the images and integrating over all intensity in the respective areas of the scattering image. We find the lamellar structure parallel to the surface of the substrate is always present, but decreasing in volume with increasing ibuprofen content. For pure SoyPC and low concentrations of ibuprofen nearly the complete volume is made up from lamellae parallel to the surface. The disordered lamellae start appearing at 13.6\,mol\% of ibuprofen and increase to a relative scattering contribution of about 0.35 at 34.5\,mol\%. They scatter around this value for all higher concentrations. All three lattices, both hexagonal and the primitive tetragonal lattice, appear simultaneously at 34.5\,mol\%. After this, at higher ibuprofen concentrations, the volume fraction of the hexagonal lattice standing on edge decreases together with the volume fraction of the primitive tetragonal lattice. This can be understood, if you regard the primitive tetragonal lattice as a filler between the two hexagonal lattices, which is necessary wherever these two meet. As the hexagonal lattice standing on edge vanishes, this filler is no more needed, and the favored hexagonal lattice with a parallel axis to the surface is the only lattice remaining at high concentrations.


\subsubsection{Neutron Reflectometry}
\begin{figure}
  \begin{minipage}[t]{0.48\textwidth}
\raggedright{\large{(a)}}
\includegraphics[width=\textwidth]{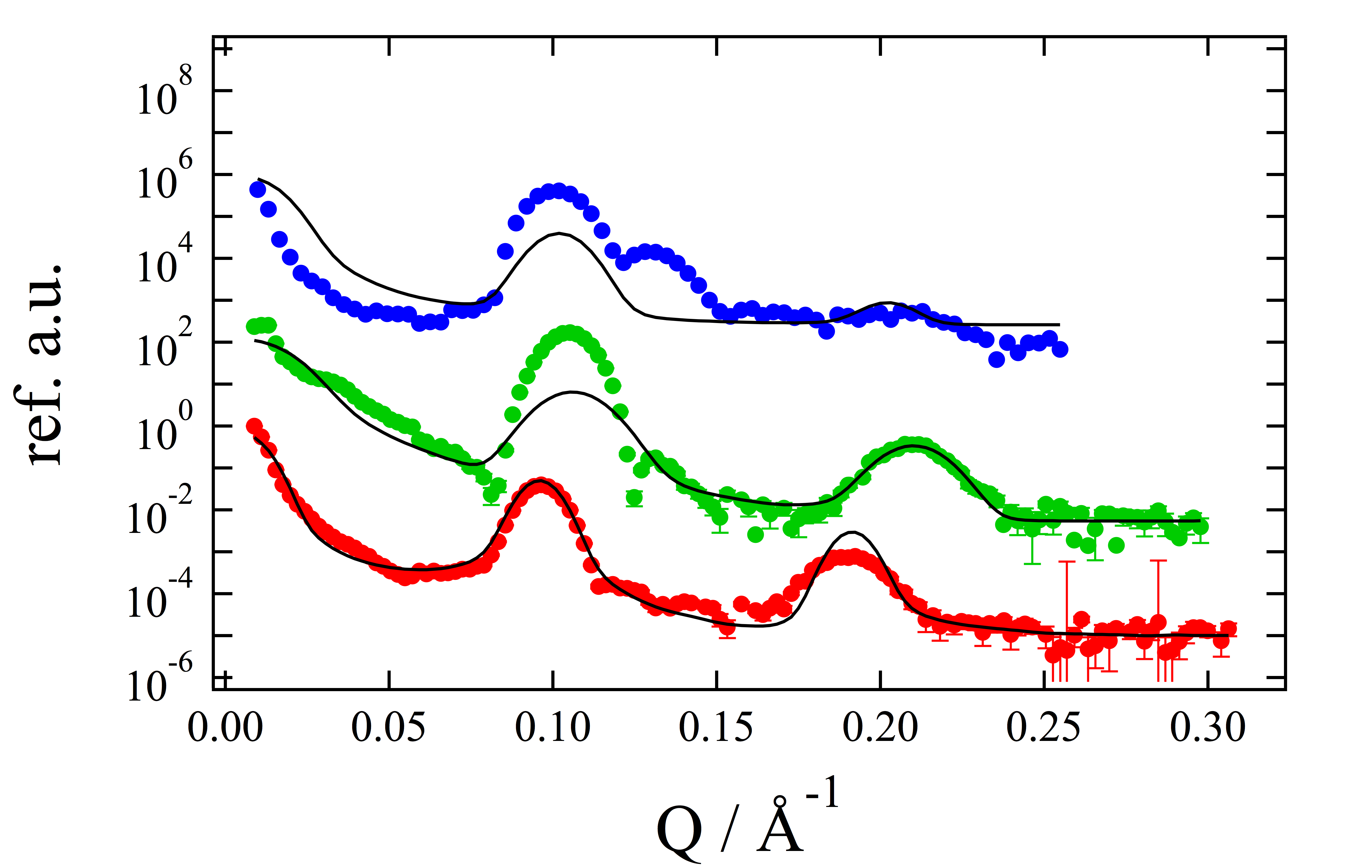}
\end{minipage}
\begin{minipage}[t]{0.48\textwidth}
\raggedright{\large{(b)}}
\includegraphics[width=\textwidth]{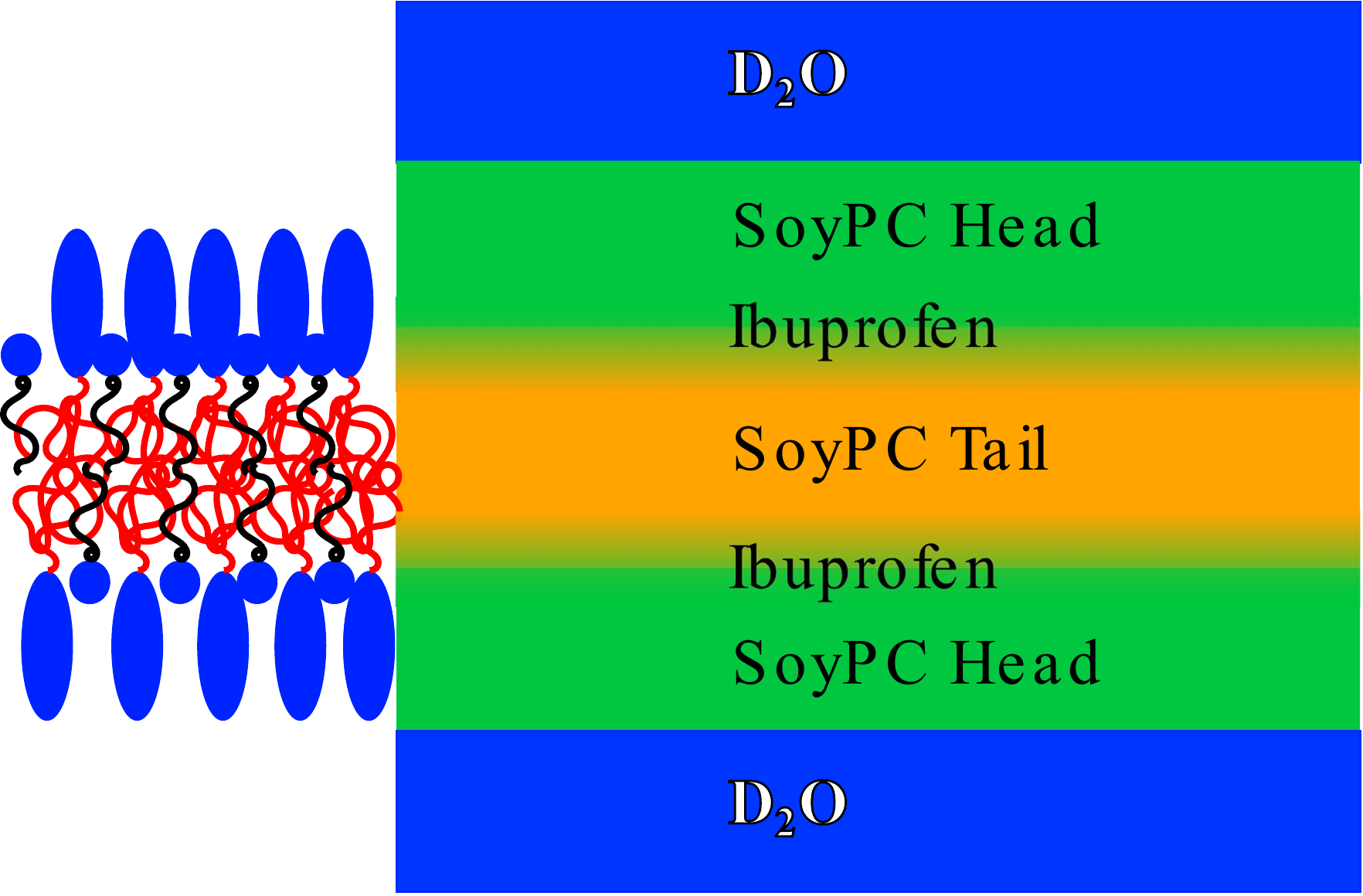}
\end{minipage}
\caption{(color) (a) Representative reflectometry data with fits. Datasets are shifted for better visibility. Concentrations of ibuprofen (bottom to top) are 0, 13.6, 25.0\,mol\%. (b) Depiction of the layer model used by the Parratt algorithm. For better visibility hydrophobic parts of SoyPC are in red, those of ibuprofen in black. \label{fig:sampleReflecometry}}
\end{figure}

Representative results with fits according to the Parratt algorithm are shown in fig.\,\ref{fig:sampleReflecometry}\,a). All major features of the reflectivity curves could be reproduced. The initially assumed distribution of the layers is depicted in fig.\,\ref{fig:sampleReflecometry}\,b). Additionally to this repeating unit, a water layer directly on the substrate with a thickness of $\approx50$\,\AA\, was assumed for all concentrations. In all systems about 35-40 repetitions of these layers were found. The existence of this water layer was corroborated by the fits. The critical angle vanishes in these systems. MARIA uses an elliptically curved focusing guide on the vertical direction, which ensures a vertical beam size of about 1\,cm on the sample position. In the horizontal direction the opening of the two collimation slits (S1,S2) that are 4\,m apart, was equal to 1\,mm for S1 and 1\,mm S2 resulting in a collimated beam of  0.5 mrad ($\theta=2\tan^{-1}(\left[S1+S2\right]/2L)$).

Discrepancies between the fit, especially in the width of the peaks, and the data can be explained by the fact that at higher concentrations, instead of investigating a purely lamellar system, which is ordered parallel to the substrate surface, additional ordering occurs. If one considers the size of the footprint of approximately ($\approx 12\mathrm{\,cm}\times 1.6\mathrm{\,cm}=19.2\mathrm{\,cm}^2$) it becomes apparent that the reflected intensity is comprised of intensity reflected from a lamellar structure as well as the ordered lattice structures. As a perfectly parallel lamellar stack is assumed in the Parratt model, concentrations above 25.0\,mol\% are not accurately described anymore and already at 25.0\,mol\% the fit is already challenging. However we still performed the same analysis of the reflectometry for all concentrations in order to obtain information about the layer structure itself, which is embedded in the sample for all concentrations. In order to account for the smaller side maxima, which we attribute to the emerging 3D structure, we chose a phenomenological approach and fitted weighted Gaussians to the peaks. Results of these fits reveal the smaller peaks to increase in relative intensity from about $10^{-3}$ at 13.6\,mol\%  to $0.1$ at 43.1\,mol\%. This is consistent with the idea, that the emerging 3D structure is induced by the ibuprofen and thus the relative scattering contribution of the attributed peak increases.

The values for the SLDs of the different layers are given in table \ref{tab:SLD}. Another problem in the accurate description may be that the layers may have an initial disorder already at very low concentrations, that is not accounted for by our analysis.

 \begin{table}
 \caption{SLDs used for the fitting of the reflectometry data with the Parratt algorithm. All values except for SoyPC are calculated using tabulated values published by NIST \cite{NIST}. SLD for SoyPC was determined using contrast variation analysis with D$_2$O/H$_2$O mixtures. \label{tab:SLD}}
 \begin{ruledtabular}
 \begin{tabular}{ccc l c|l ccc}
& & & Component&	&SLD [$10^{-6}$\AA$^{-2}$]& & &	\\
& & & Silicon&		&2.08& & &	\\
& & & D$_2$O	&	&6.38& & &	\\
& & & SoyPC&		&0.24& & &	\\
& & & Decane&		&-0.49& & &	\\
& & & Ibuprofen&	&0.92& & &	\\
 \end{tabular}
 \end{ruledtabular}
 \end{table}

\begin{figure}
\begin{minipage}[t]{0.48\textwidth}
\raggedright{\large{(a)}}
 \includegraphics[width=\textwidth]{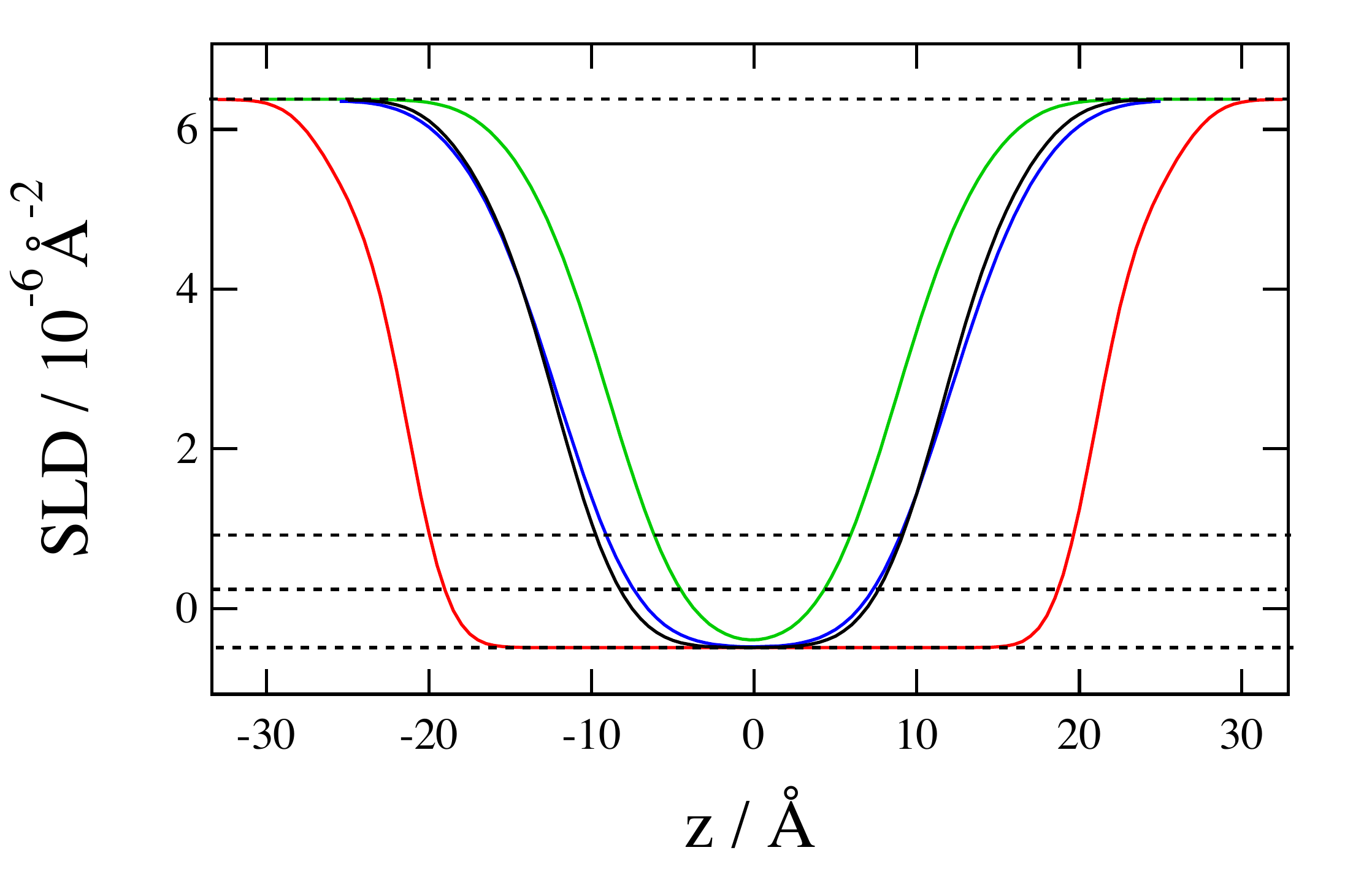}
\end{minipage} 
\begin{minipage}[t]{0.48\textwidth}
\raggedright{\large{(b)}}
 \includegraphics[width=\textwidth]{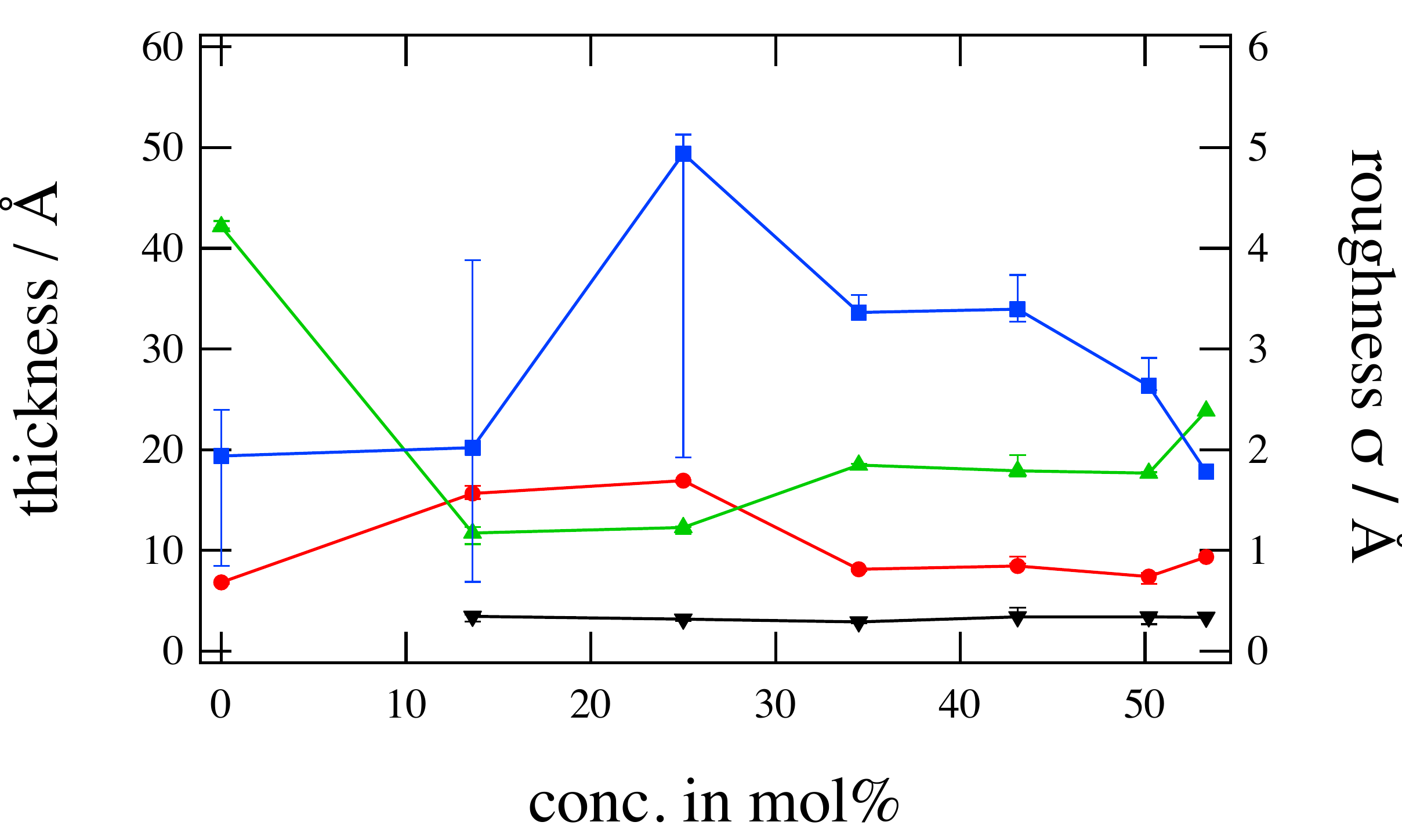}%
\end{minipage}
 \caption{(color online) (a) SLD profiles from fits with the Parratt algorithm to reflectometry data for all concentrations. Concentrations are 0\,mol\% (red), 13.6\,mol\% (green), 43.1\,mol\% (blue), and 53.3\,mol\% (black). Dashed lines indicate the SLD from table\,\ref{tab:SLD} for decane, SoyPC, ibuprofen and D$_2$O (from bottom to top), (b)\,results from modelling the reflectometry curves with the model shown in fig. \ref{fig:sampleReflecometry}: Thickness of the D$_2$O layer (red circles), thickness of the SoyPC layer (green upright triangles), roughness of the layers (blue squares) and thickness of the ibuprofen layer (black downright triangles). \label{fig:SLD-profiles}}
 \end{figure}

Calculated SLD-profiles from eq. \ref{eq:ref-index} are shown in fig. \ref{fig:SLD-profiles} a). These profiles show that for the pure SoyPC layers the pure hydrocarbon layer, which was used to model the tail region of SoyPC is about 40\,\AA\, wide(for comparison of the respective SLDs used see table \ref{tab:SLD}). After the introduction of ibuprofen into the system this width immediately collapses down to only about 20\,\AA\, in thickness. We attribute this apparent collapse to the ibuprofen becoming an interstitial part of the membrane and thus changing the overall SLD profile at the interface. With an increasing amount of ibuprofen the thickness of the SoyPC layer consecutively increases again, but does not reach its initial thickness anymore.

The different thicknesses for all layers used in this model can be seen in fig. \ref{fig:SLD-profiles} b). The sudden decrease which is found in the SLD profiles is reproduced. At first glance it is surprising, that the thickness of the ibuprofen layer should be constant for all concentrations, considering the concentration is increased from 13.6\,mol\% to 53.3\,mol\%. However, taking into account the increasing thickness of the SoyPC layer, which is primarily the thickness of the hydrocarbon tail, we see a steady increase along with increasing concentration. This can be explained by assuming the ibuprofen is preferentially dissolved by the hydrocarbon tail of the SoyPC and thus inflates this layer.

Another observation from this data is, that the modelled roughness of the layers is maximal with about 5\,\AA\, at 25.0\,mol\% ibuprofen. We assume this roughness is correlated to a high strain of the membrane which occurs at the onset of ordering to accommodate for different conformations within the layer stack. As the concentration is increased the ordering is improving again, so the roughness of the layers decreases.

\subsubsection{GINSES}
\begin{figure}
 \includegraphics[width=0.8\textwidth]{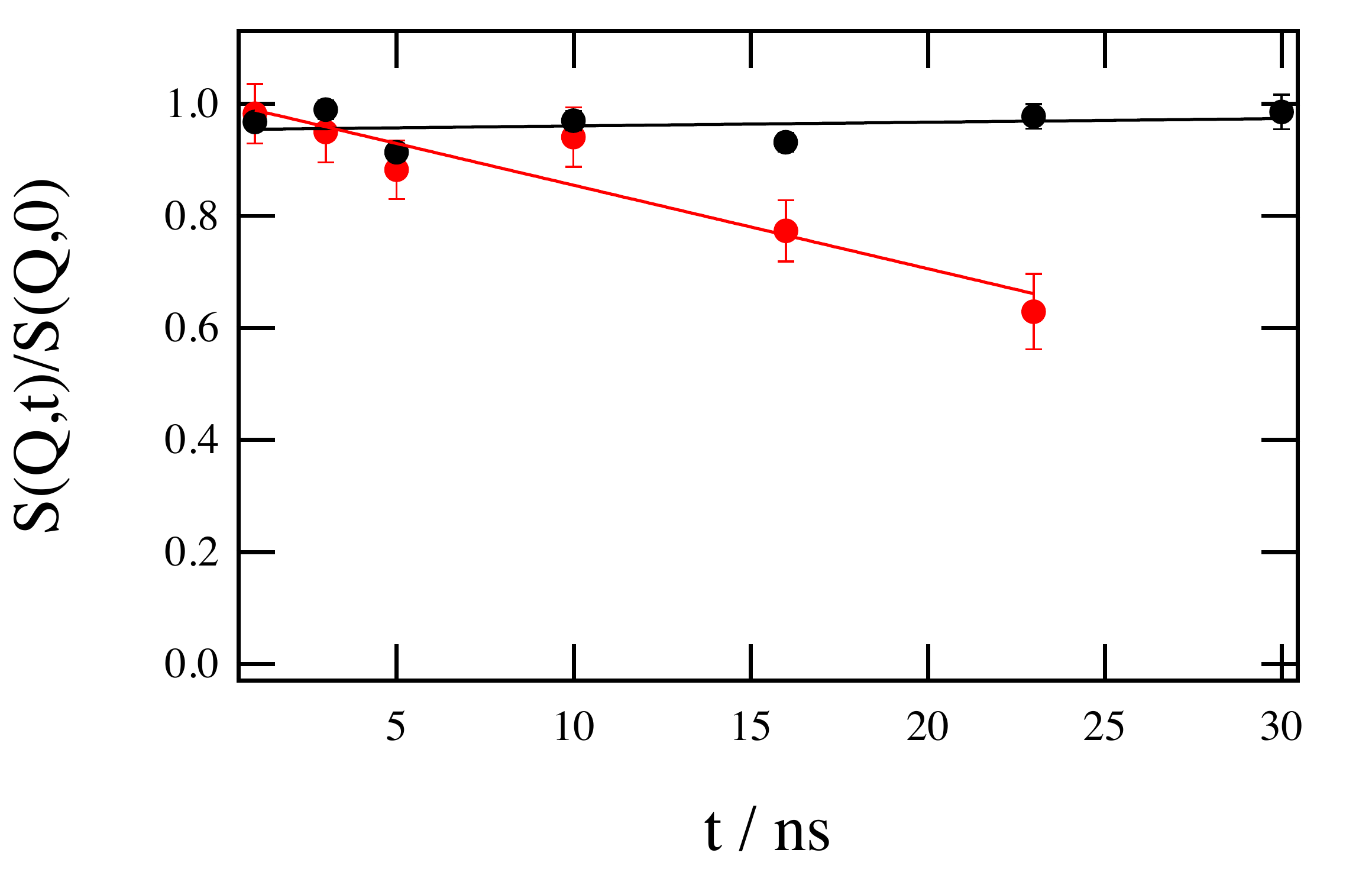}%
 \caption{(color online) Dynamic structure factors for 0 (red) and 34.5\,mol\% (black) of ibuprofen in a SoyPC layer. For the pure SoyPC a distinct relaxation in the investigated time regime is visible, while for 34.5\,mol\% the sample is does not show any relaxations.\label{fig:GINSES_results}}
 \end{figure}

The results from the GINSES measurements are shown in fig. \ref{fig:GINSES_results}. Due to the long measurement times only the samples with pure SoyPC and the sample with 34.5\,mol\% of ibuprofen were investigated. While a relaxation is visible for the pure SoyPC, the sample with additional ibuprofen shows no relaxation. There are too few points to fit a meaningful relaxation time, however this behavior can qualitatively be interpreted as a stiffening of the membranes with increasing ibuprofen content. The measurements were performed for $Q=0.12$\,\AA$^{-1}$, which translates to an evanescent wave depth of $\Lambda_{\mathrm{eva}}\approx380\mathrm{\,\AA}$.

The specific choice for the Q-value can be rationalized by the fact, that at correlation peaks the dynamics get very slow due to the so-called deGennes narrowing, where the relaxation rate is proportional to the inverse form factor \cite{deGennes1959,Holderer2007}.

Between Bragg peaks, in a minimum of the static structure factor, the signature of dynamic fluctuations on the form factor of the membranes is better visible, it is therefore an advantage to measure membrane fluctuations between Bragg peaks.

Local fluctuations of membranes, such as contrast between water and double layer, or density fluctuations of the scattering length density in the membrane, can result in a relaxation of the intermediate scattering function. This will be visible at high Q, and preferably not at structure factor peaks.

These local fluctuations of a double layer are visible in the form factor, not the structure factor, which makes the dynamics of different compositions comparable, even if the structure differs.

\section{Discussion}
The introduction of ibuprofen into SoyPC phospholipid films has several effects, which are related to one another. The thinning of the lipid film upon the introduction of ibuprofen, as well as the stiffening coincide with first a breaking up of the parallel lamellae and, then with the emergence of several coexisting lattices and finally with a single hexagonal lattice at 53.3\,mol\% ibuprofen in the film.

Following this behavior step by step it is possible to connect all these behaviors: (1) Evolution of a lamellar powder, (2) emergence of several lattices, (3) thinning of the SoyPC layers and (4) stiffening of the surface.

\begin{figure}
  \begin{minipage}[b]{0.45\textwidth}
    \raggedright{\large{(a)}}
  \includegraphics[width=\textwidth]{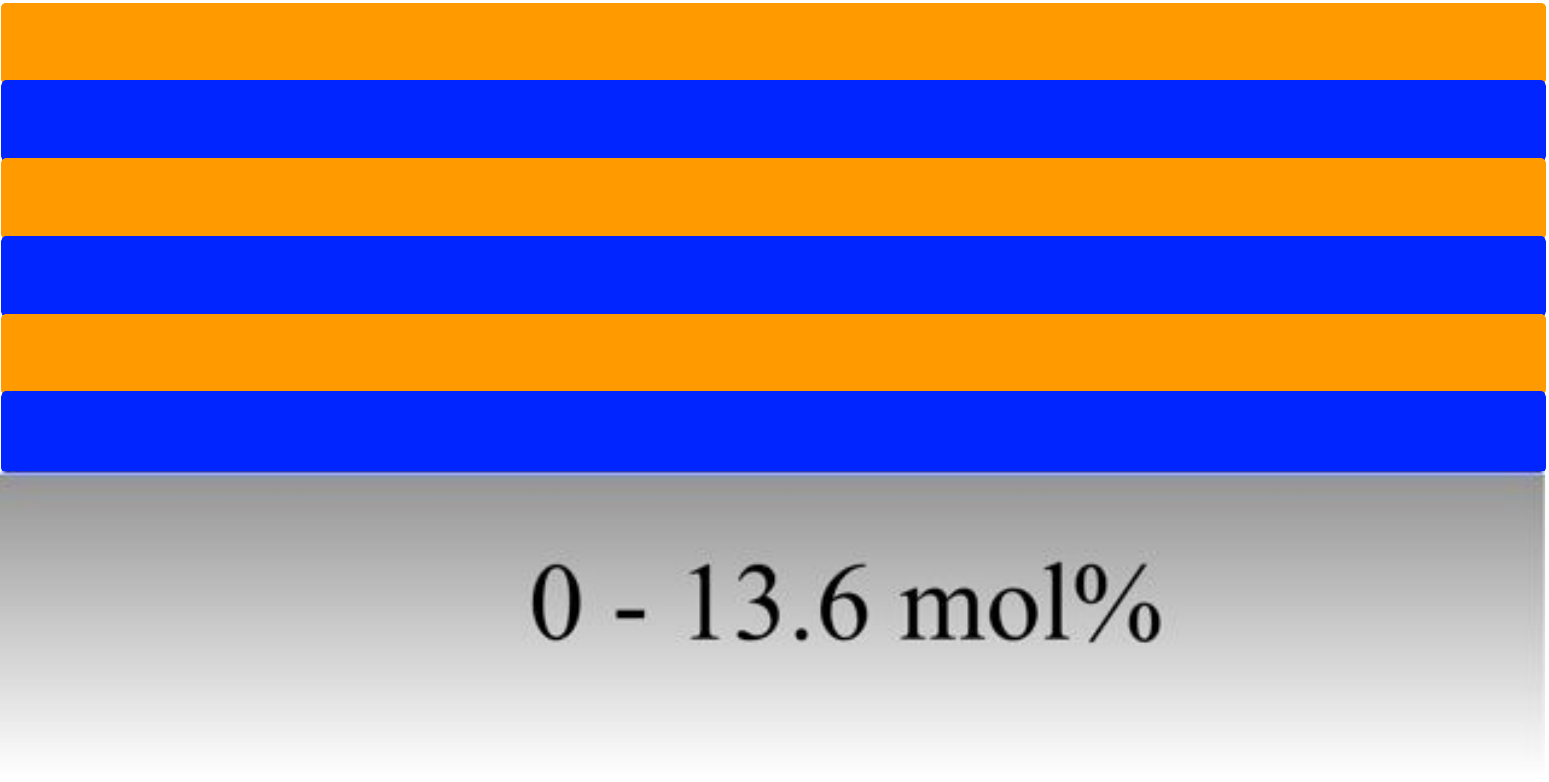} 
  \end{minipage}
  \begin{minipage}[b]{0.45\textwidth}
    \raggedright{\large{(b)}}
  \includegraphics[width=\textwidth]{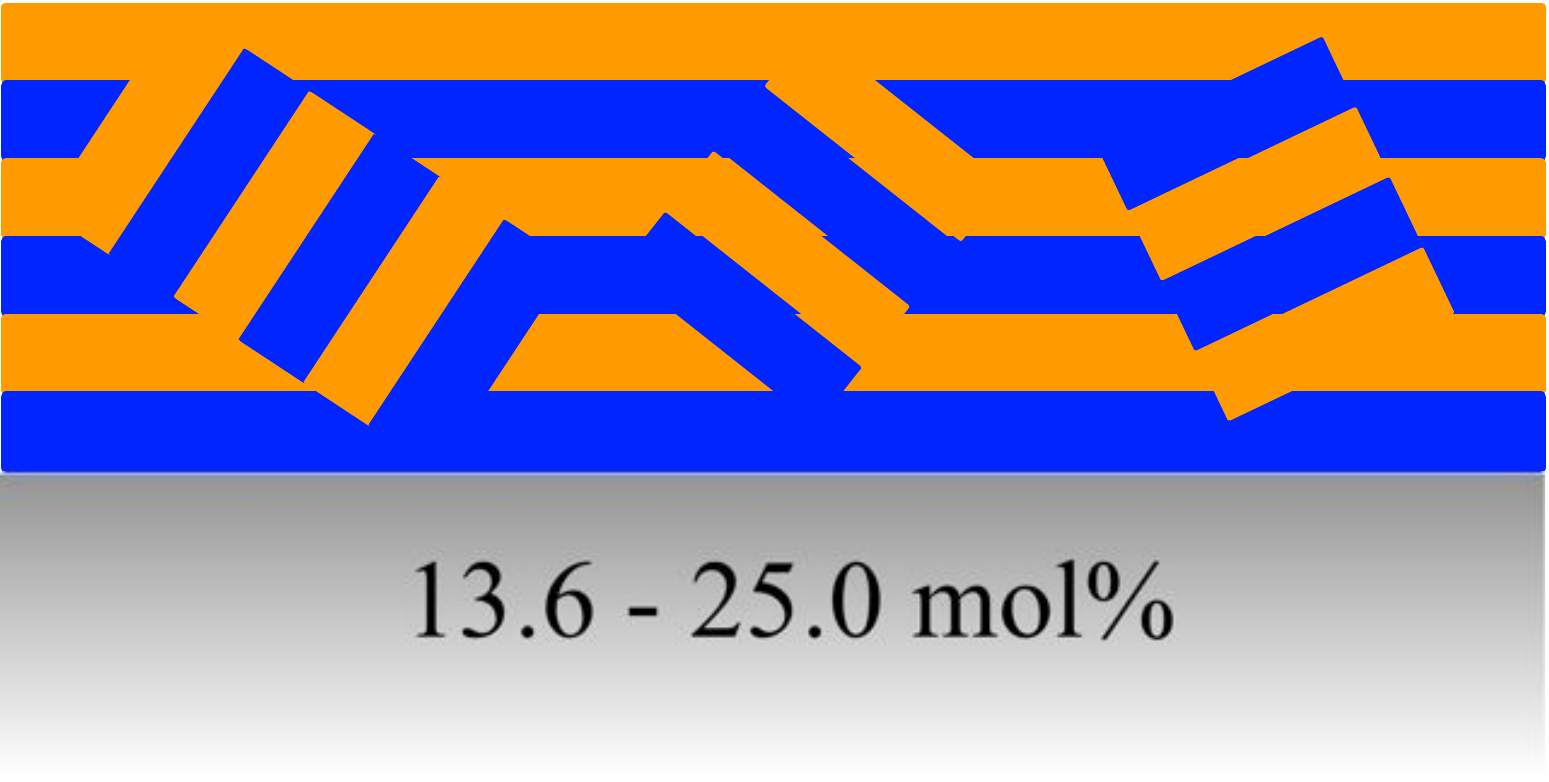} 
  \end{minipage}
  
  \begin{minipage}[b]{0.45\textwidth}
    \raggedright{\large{(c)}}
  \includegraphics[width=\textwidth , height=0.8\textwidth]{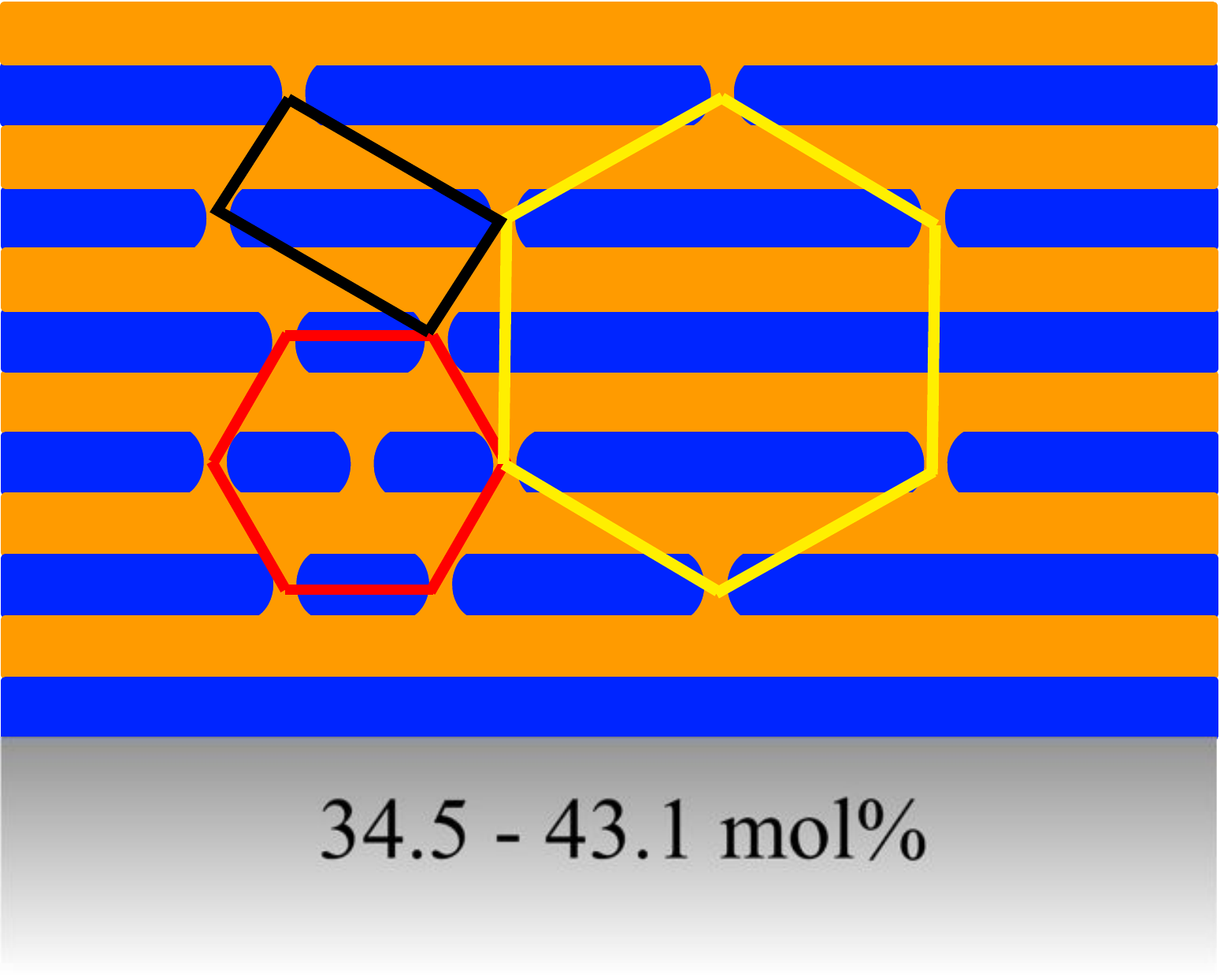} 
  \end{minipage}
  \begin{minipage}[b]{0.45\textwidth}
    \raggedright{\large{(d)}}
  \includegraphics[width=\textwidth , height=0.8\textwidth]{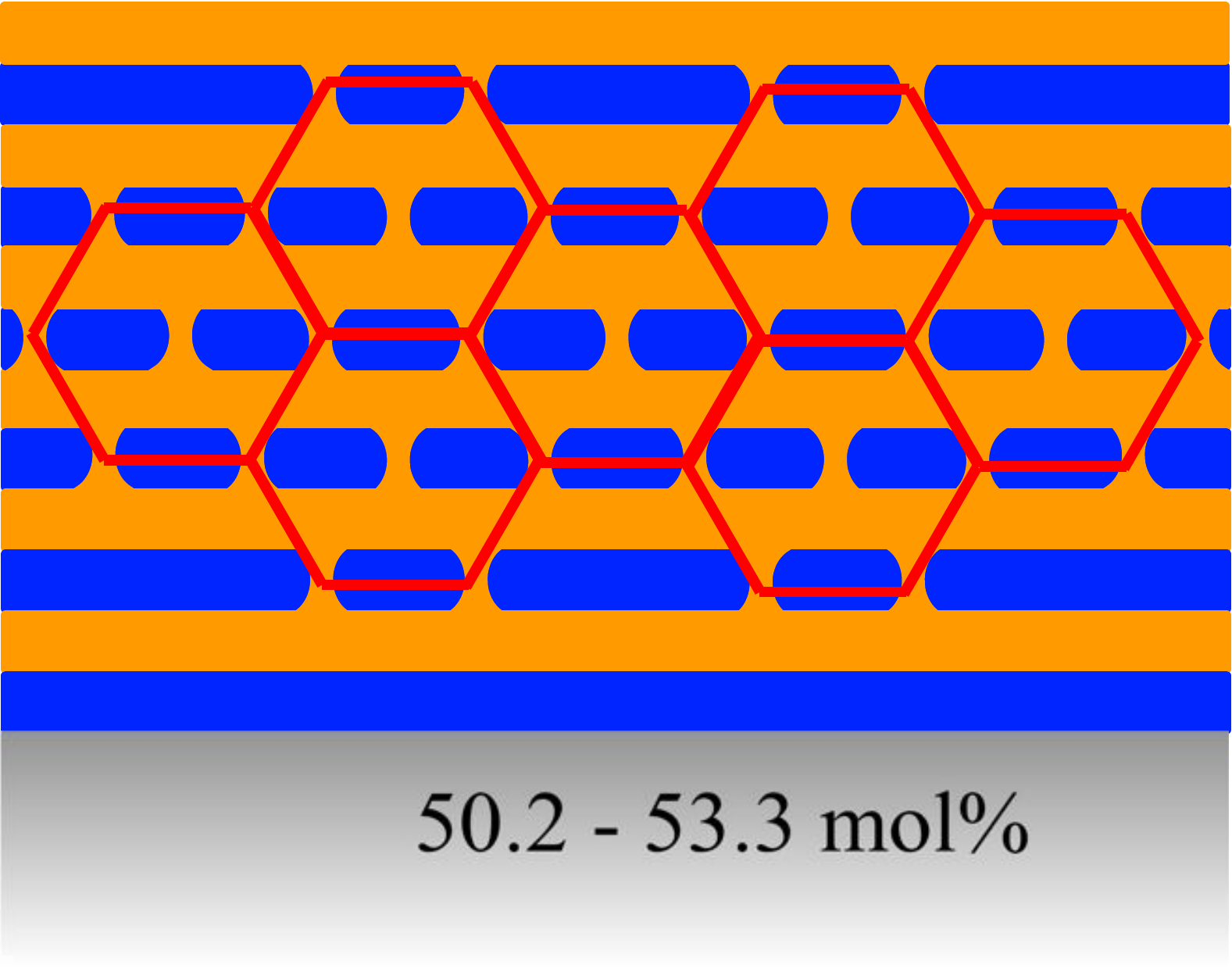} 
  \end{minipage}
   \caption{(color online) Sketch of the structural evolution of the sample. (a) For low concentrations of ibuprofen the system is dominated by a lamellar structure while introduction of more ibuprofen (b) induces disordering of lamellar areas and thus powder scattering of lamellar areas. In an intermediate concentration area (c) there are two different hexagonal lattices which are stabilized against each other by a primitive tetragonal lattice. At very high concentrations (d) only the hexagonal structure with an axis parallel to the substrate is retained. Color coding of the hexagonal structure corresponds to the color coding used in Fig.\,\ref{fig:GISANS-images}. \label{fig:structure-evolution}}
 \end{figure}

(3) and (4) are connected, assuming that the introduction of ibuprofen is indeed similar to the drying of the lamellae, which is supported by the comparison with data from Aeffner et al. \cite{Aeffner2012}, where a similar behavior was found for the drying of a phospholipid film. This drying in turn leads to a strain in the surface, as a dry film becomes less flexible and is less apt to follow the zero curvature of the substrate surface, but will prefer a curvature which is determined by the packing parameter of the SoyPC. The evolution of a lamellar powder (1) is also a result of this. As the strain on the surface increases and the curvature is more and more determined by a very stiff surface with a high curvature, lattices form (2) in order to accommodate for this high curvature of the different lamellae. Finally, when the strain is high enough and the lamellae are very stiff, there is only one possible conformation of a lattice that can accommodate for this high strain. This behavior is sketched in fig. \ref{fig:structure-evolution}. In each of the panels the newly emerging structure is highlighted, but others, such as the disordered lamellae may still be present (see fig. \ref{fig:rel_intensity}). It is striking that in this representation it is not possible to create hexagonal lattices with equal spacings for both orientations, which might be suggested looking at the GISANS images. There are two possible approaches to that: (1) The Q-space resolution of the GISANS images is not able to resolve this difference in lattice spacing, which amounts to $4D_{lam}$ for the hexagonal lattice standing on edge versus $3D_{lam}\sin 60^\circ \approx 2.6 D_{lam}$. This ratio of $4/2.6\approx 1.5$ translates into a similar difference in Q-spacing. Looking at the GISANS images in fig. \ref{fig:GISANS-images} it is conceivable that a factor of 1.5 is not visible between the two different hexagonal lattices, as the maxima are quite broad. (2) It is possible that there are different regions where the lamellar thickness is not constant and at the same time one or the other hexagonal lattice may be predominant.

Regarding the inclination of the primitive tetragonal lattice we could confirm via comparisons of scattering images created with the software XrayView 4.0 that the lattice is indeed inclined by $\approx 75^\circ$. This compares to an inclination of $\approx 60^\circ$ when regarding the conformation as envisioned in fig. \ref{fig:structure-evolution}\,c. While this deviation is substantial, considering the low volume fraction and thus the low intensity scattered from the primitive tetragonal lattice, which accommodates both hexagonal lattices, it still seems a good fit.

\begin{figure}
  \begin{minipage}[t]{0.4\textwidth}
    \raggedright{\large{(a)}}
  \includegraphics[width=\textwidth]{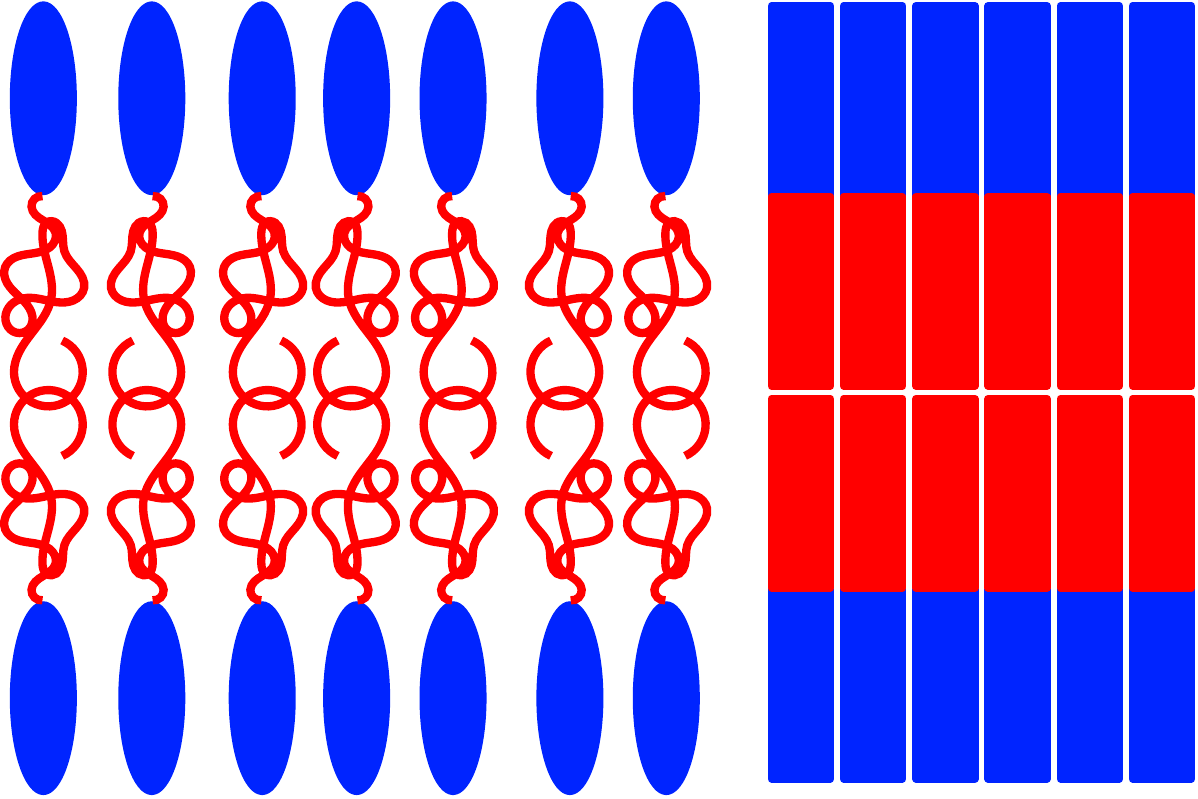} 
  \end{minipage}
  \begin{minipage}[t]{0.4\textwidth}
    \raggedright{\large{(b)}}
  \includegraphics[width=\textwidth]{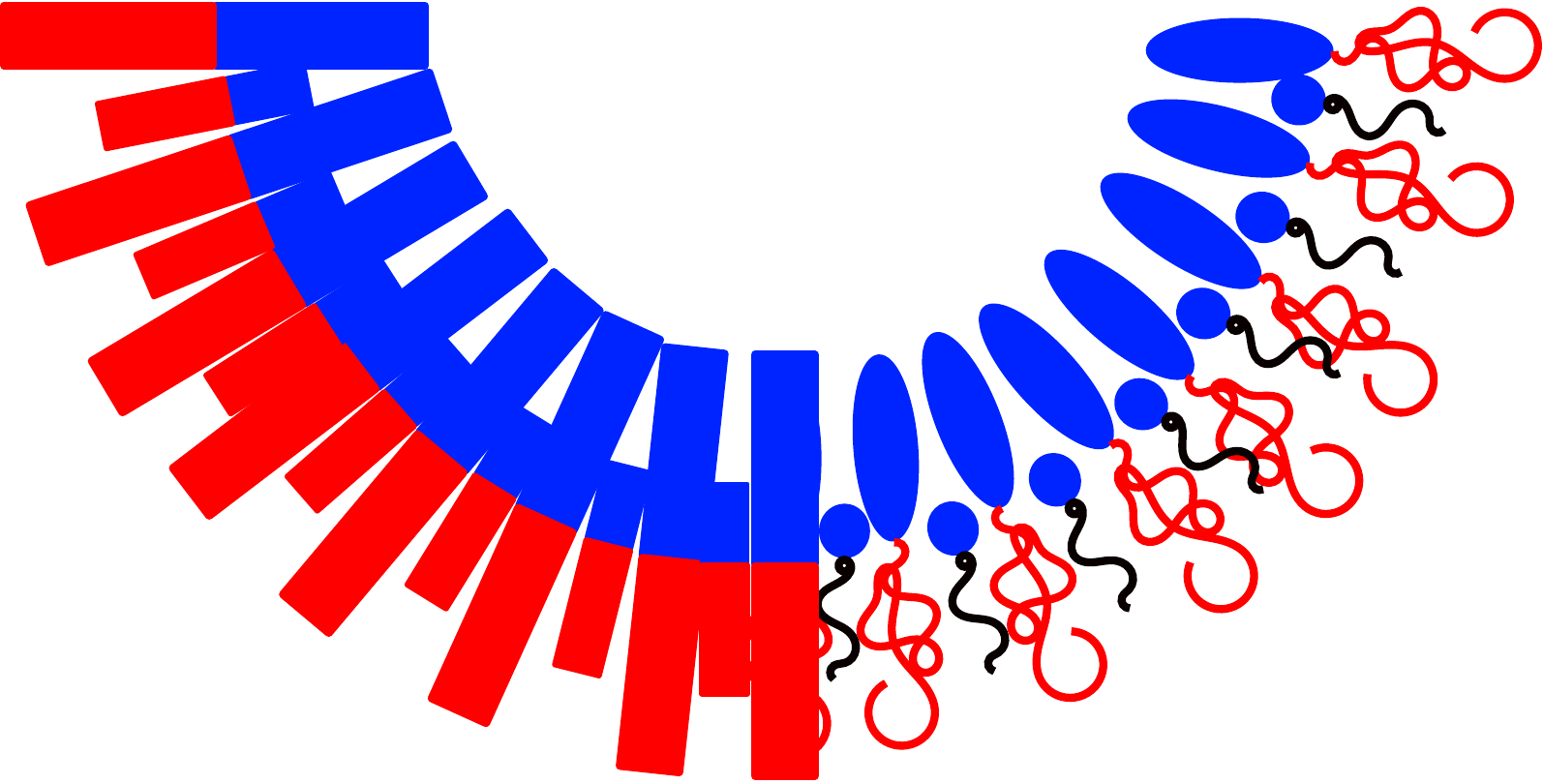} 
  \end{minipage}
   \caption{(color) Sketches of packing for a) pure SoyPC layers and b) SoyPC and ibuprofen. The ibuprofen with the small hydrophilic unit increases the curvature of the multilayer forming SoyPC. \label{fig:packingparameter}}
 \end{figure}

\begin{figure}

  \includegraphics[width=0.8\textwidth]{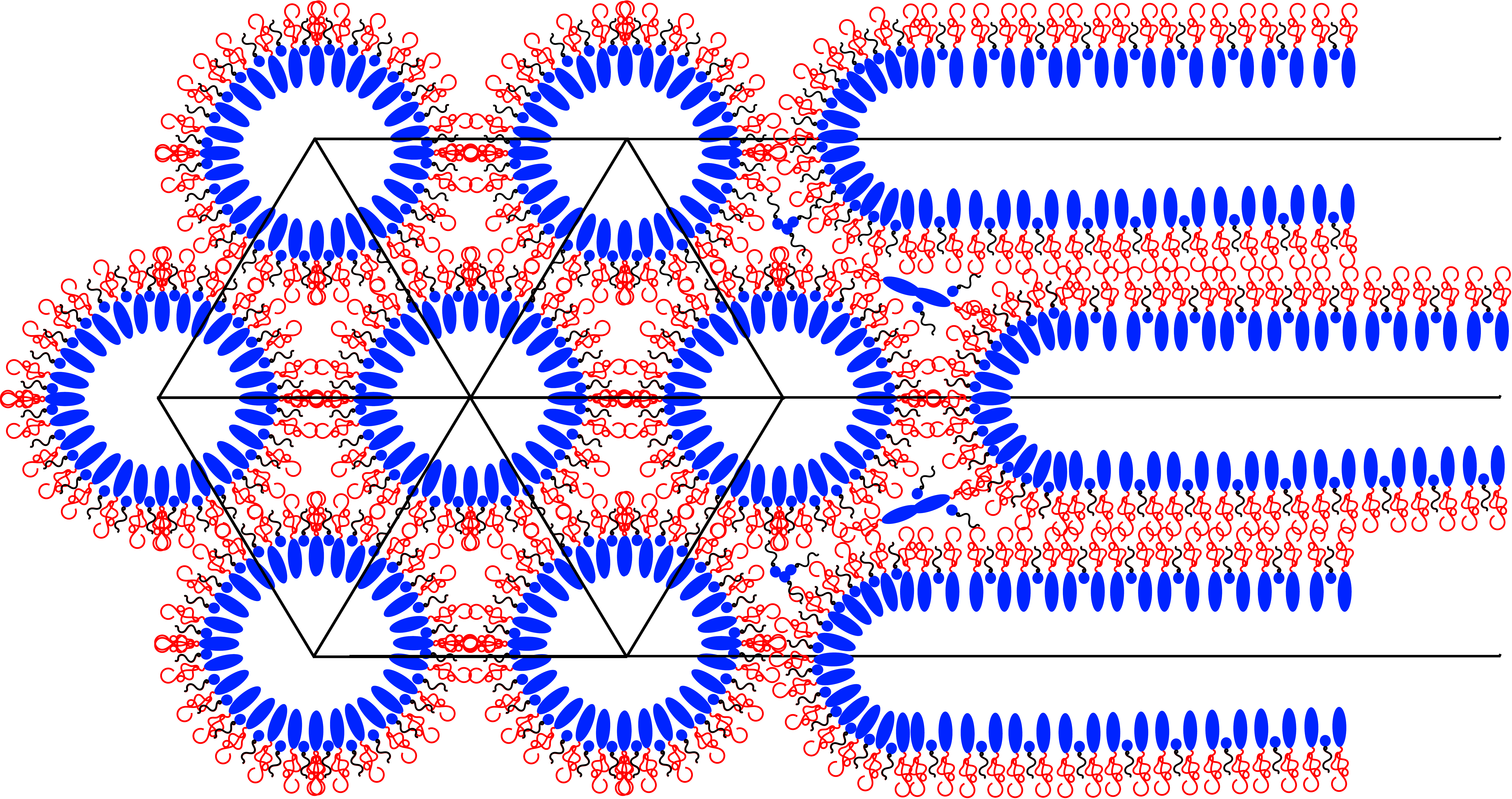} 

   \caption{(color) Sketches of hexagonal ordering induced by the introduction of ibuprofen into the SoyPC matrix. The smaller ibuprofen increases the curvature in the hexagonal phase, while the lower amount of ibuprofen in the lamellar phase allows for a more parallel structure. Here it is important to note that the concentration differences between the different phases may be very minute. However, the emergence of different phases in the sample seems to corroborate the possibility of a slightly inhomogeneous distribution.\label{fig:hexpacking}}
 \end{figure}

To explain this behavior on a molecular level, a consideration of the packing parameter as proposed by Israelachvili \cite{Israelachvili} is helpful. While SoyPC exhibits a packing parameter close to $p_{\mbox{pack}}=v/l\cdot a_0$ where $v$ is the volume of the hydrocarbon chain of the lipid in solution, $l$ is its length and $a_0$ is the surface area in an aggregate occupied the the hydrophilic part this value decreases strongly as soon as ibuprofen is introduced, as can be seen in Fig.\,\ref{fig:packingparameter}. The rational for the development of the hexagonal structure is, that the change in the average packing parameter by introducing ibuprofen into the SoyPC is lowered. This happens as the hydrophobic part of the ibuprofen is much smaller than in the case of SoyPC and thus induces a higher curvature. If the curvature is high enough, the hexagonal structure emerges, while still parts of the lamellar structure are retained as in the case of the pure SoyPC (see fig.\,\ref{fig:hexpacking}). This explanation is largely based on the location of the ibuprofen being interstitial between the head and tailgroups of the SoyPC. This assumption is corroborated by the reflectometry data as well as by computer simulations \cite{Paloncyova2013} and x-ray \cite{Barrett2012} scattering based electron density found in the literature. Although in these publications smaller molecules were investigated the physical determining features, namely small size and amphiphilicity, are identical. This seems to hint to a general attraction of small molecule drugs with amphiphilicity to the boundary between head- and tailsegment of lipids in a membrane. Geometric considerations like that of Israelachvili \cite{Israelachvili} corroborate that in this area the influence of the drugs on the bending modulus and hence the structure of the membrane is most pronounced.

\section{Conclusion}
We investigated the influence of the ibuprofen concentration on the behavior of phospholipid films of SoyPC. We found a correlation between the ibuprofen content and the conformation of the lamellae in the film, starting as a film of nearly perfectly parallel lamellae when still a pure SoyPC film, then at low concentrations of ibuprofen exhibiting powder scattering of disordered lamellae. In an intermediate state there is a coexistence of two hexagonal lattices, one parallel to the substrate surface, one standing on edge which are both stabilized by a low amount of a inclined primitive tetragonal lattice. At high concentrations only the hexagonal lattice parallel to the substrate and the lamellae, both parallel and disordered, are retained. 

The ordering of the lipid layers shown by the GISANS measurements explain the damage done by ibuprofen to cell membranes. Any membrane that undergoes an ordering will not constitute a completely closed membrane anymore. Considering the SoyPC as a physical model system this damage to the membrane is consistent with the stomach bleeding found in the case of long exposure

Along another line of thought there seem to be a variety of possibilities how potentially non-dangerous changes of parameters can induce ordering, and thus destroy, lipid membranes. Considering the induced order by drying as investigated by Aeffner et al. \citep{Aeffner2012} there seem to be parallels which will have to be investigated further.

However, here we have to keep in mind that the ibuprofen concentrations investigated here are beyond any medical applicability, so this same effect cannot be expected in medical practice. It is however conceivable, that in the case of long-term treatment, where these complications occur, once a \textit{nucleation point} for this damage is created, the damage will start to grow. This initial damage can be due to a local, short-time high concentration immediately after ingestion. Here we want to stress that for the nucleation point we are strictly speaking in terms of likely-hood. This means, the individual nucleation point does not need to be stable over a long time, but that during a frequent exposure to high doses of ibuprofen the probability for the formation of such a nucleation point is strongly increased.

Apart from the structural damage, a structure induced by ibuprofen and the stiffening in itself may alter the mobility of proteins in the membrane before damaging it. However this change in mobility may inhibit the protein function thus damaging the cell nonetheless \cite{Saxton1987}.

\bibliography{apssamp}

\newpage

\begin{center}
\begin{large}
\textbf{SI: Influence of Ibuprofen on Phospholipid Membranes}
\end{large}
\vspace{1cm}

Sebastian Jaksch, Frederik Lipfert, Alexandros Koutsioubas, Stefan Mattauch, Olaf Holderer, Oxana Ivanova, Henrich Frielinghaus

\textit{Forschungszentrum J\"ulich GmbH, JCNS at Heinz Maier-Leibnitz Zentrum, Lichtenbergstra\ss e 1, D-85747 Garching}
\vspace{1cm}

Samira Hertrich, Stefan F. Fischer, Bert Nickel

\textit{Ludwig-Maximilians-Universit\"at, Department f\"ur Physik und CeNS, Geschwister-Scholl-Platz 1, D-80539 M\"unchen}

(Dated: \today)
\end{center}
\vspace{1cm}

  \begin{figure}[h]

  \includegraphics[width=\textwidth]{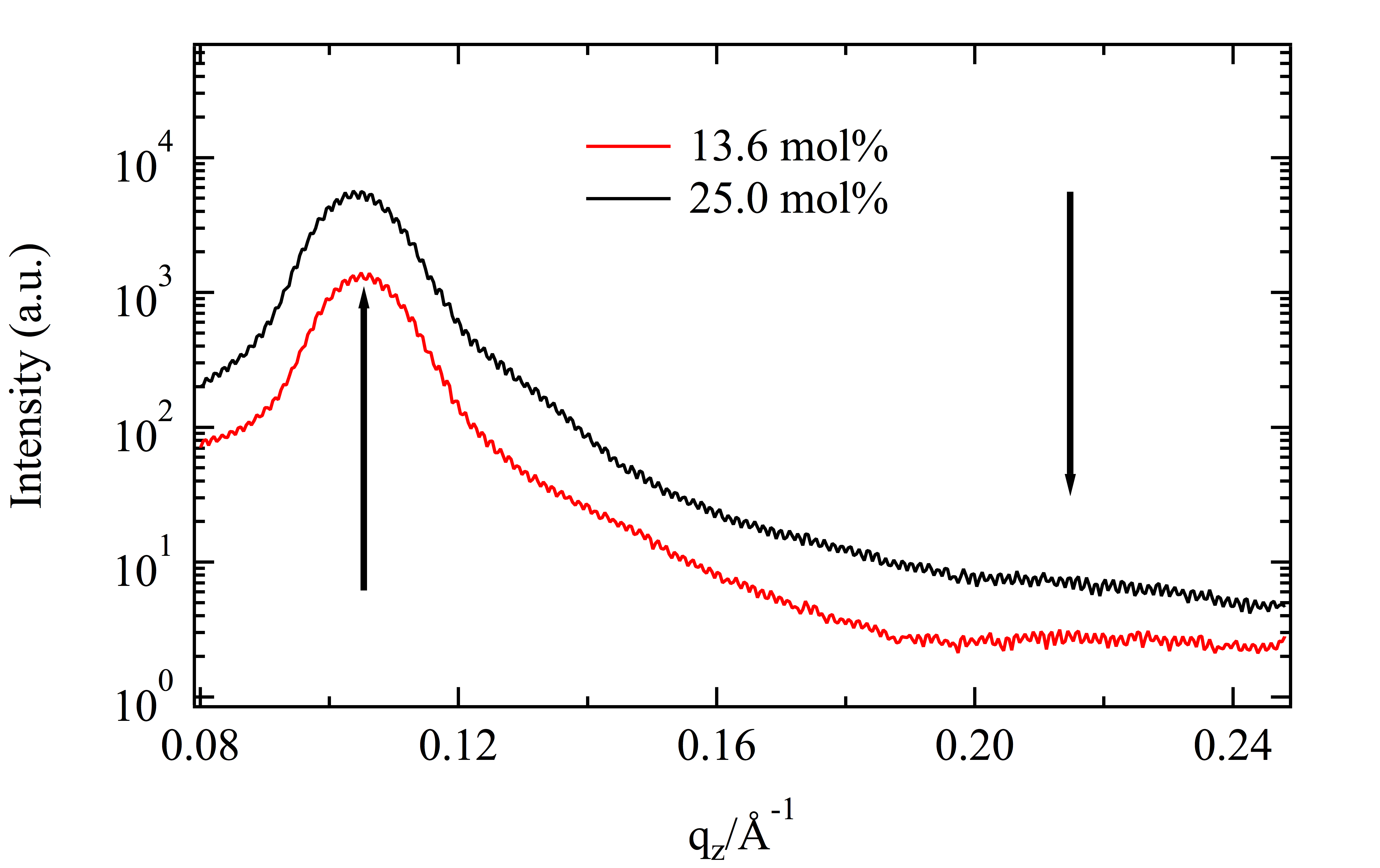} 

   \caption{The second maximum is not visible in the GISANS images with the bare eye. In the linecut the second maximum of the peak at $q_Z = 0.105$\AA$^{-1}$ is just discernible as a shoulder.}
 \end{figure}
\end{document}